\documentclass{article}

\pdfoutput=1
\usepackage{arxiv}
\usepackage[numbers]{natbib}
\usepackage[utf8]{inputenc} 
\usepackage[T1]{fontenc}    
\usepackage{hyperref}       
\usepackage{url}            
\usepackage{booktabs}       
\usepackage{amsfonts}       
\usepackage{nicefrac}       
\usepackage{microtype}      
\usepackage{lipsum}		
\usepackage{graphicx}

\usepackage{doi}

\usepackage{pgfplots}
\pgfplotsset{compat=1.8}
\usetikzlibrary{
        pgfplots.dateplot,
    }
\usepackage{subcaption}
\usepackage{algorithmic}
\usepackage{amsmath}
\usepackage{amsfonts}
\usepackage{algorithm}
\usepackage{amsthm}
\usepackage{amssymb}
\usepackage{graphicx}
\usepackage{diagbox}

\title{Local Learning at the Network Edge for Efficient \&  Secure Real-Time Predictive Analytics}

\author{ \href{https://orcid.org/0000-0002-6041-2221}{\includegraphics[scale=0.06]{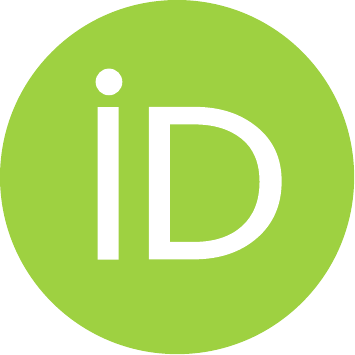}\hspace{1mm}Natascha Harth}\\
	BMW Group Research\\
	New Technologies, Innovations\\
	Parkring 19, 85748 Garching, Germany\\
	\texttt{natascha.harth@bmwgroup.com} \\
	\And
	\href{https://orcid.org/0000-0002-8401-2003}{\includegraphics[scale=0.06]{orcid.pdf}\hspace{1mm}Hans-Joerg Voegel} \\
	BMW Group Research\\
	New Technologies, Innovations\\
	Parkring 19, 85748 Garching, Germany\\
	\texttt{hans-joerg.voegel@bmwgroup.com} \\
	\And
	Kostas Kolomvatsos \\
	Department of Computer Science and Telecommunications\\
	University of Thessaly\\
	\texttt{kostasks@uth.gr} \\
	\And
	\href{https://orcid.org/0000-0003-1517-6757}{\includegraphics[scale=0.06]{orcid.pdf}\hspace{1mm}Christos Anagnostopoulos} \\
	School of Computing Science\\
	University of Glasgow 
	\texttt{christos.anagnostopoulos@glasgow.ac.uk} \\
}

\date{}

\renewcommand{\headeright}{Harth et al.}
\renewcommand{\undertitle}{ }
\renewcommand{\shorttitle}{ }

\hypersetup{
pdftitle={Local Learning at the Network Edge for Efficient \&  Secure Real-Time Predictive Analytics},
pdfauthor={Natascha Harth et al.},
pdfkeywords={Federated Learning, Quality-Aware, Personalization, Optimal Stopping Theory},
}

\begin{document}
\maketitle

\begin{abstract}
The ability to perform computation on devices, such as smartphones, cars, or other nodes present at the Internet of Things leads to constraints regarding bandwidth, storage, and energy, as most of these devices are mobile and operate on batteries. Using their computational power to perform locally machine learning and analytics tasks can enable accurate and real-time predictions at the network edge. A trained machine learning model requires high accuracy towards the prediction outcome, as wrong decisions can lead to negative consequences on the efficient conclusion of applications. Most of the data sensed in these devices are contextual and personal requiring privacy-preserving without their distribution over the network. When working with these privacy-preserving data, not only the protection is important but, also, the model needs the ability to adapt to regular occurring concept drifts and data distribution changes to guarantee a high accuracy of the prediction outcome. 
We address the importance of personalization and generalization in edge devices to adapt to data distribution updates over continuously evolving environments. The methodology we propose relies on the principles of Federated Learning and Optimal Stopping Theory extended with a personalization component. 
The privacy-efficient and quality-awareness of personalization and generalization is the overarching aim of this work.  
\end{abstract}

\section{Introduction}

\subsection{Motivation \& Challenges}

Measurements of the surrounding environment and the continuous creation of data are becoming pervasive in our daily lives. Most of these devices are connected over the internet to transfer the data to a Central Location (CL), e.g., the cloud, for further analysis 
and processing. This collected data contains user and personal information that needs protection. Privacy 
exhibits an increased importance over the 
recent years with regulations 
defined for such purposes, such as General Data Protection Regulation (GDPR) \cite{REGULATIONSRegulation} or California Consumer Privacy Act (CCPA) \cite{CaliforniaAB-375}. Companies are forced to collect fewer and process less sensitive data of users. Consequently, sharing raw data over the network to train and deploy high-quality Machine Learning (ML) algorithms is constrained by privacy-preserving regulations and the concerns of users to share the data. 

The tremendous increase of devices raises bottlenecks of transmission and data collection due to limited bandwidth, storage, and connectivity losses. The  
Internet of Things (IoT) 
extended over time their nature from simple sensors into small computers with the ability to perform computation at the source of the data 
to overcome the bottlenecks of 
legacy 
centralized architectures. 
We can imagine an infrastructure `covering' the IoT 
i.e.,
Edge Computing (EC) 
which uses this computational capacity to enable ML at the devices, raising the possibility to achieve real-time local predictive analytics and actuation due to significantly reduced latency. However, 
EC deploys a central training of the ML model and only implementing inference at the edge. In the period of training the centralized ML algorithm transfers raw data from edge devices to a central collection point. This results again in privacy concerns and issues towards the collected and transferred data. Therefore, not only is the privacy of data the motivation for pushing intelligence and computation to the devices but also to overcome bottlenecks of bandwidth and real-time decision making. Federated Learning (FL) is an upcoming methodology to generate knowledge from data without sharing the actual raw data \cite{Konecny2016}. The aim is to create a global model by performing the learning at the Edge Devices (EDs) and transmitting only the locally updated ML model parameters over the network. This allows distributing the model's training to the device level and keeping the data private. It enables to build general knowledge at a central location without revealing any individual device information, data or context. FL also provides a solution to overcome the introduced constraints by transmitting only models over the network, which reduces the bandwidth. Further FL enables the devices to perform analytics at the device without waiting for central decision making, which reduces and eliminates any latency of the analytical outcome.    

A major challenge in ML is that many applications rely on time series and evolving data, such as contextual data streams for prediction and analysis. This data is, by its nature, changing over time and usually involves 
sudden 
updates in the
underlying 
distribution and behaviors. These concept drifts require continuous re-training of a developed model. Centralized online learning techniques have been presented mainly using Gradient Descent to build a generalized model over the data, which is evolving. Primarily, sliding window methodologies have been used to consider qualitative analysis over evolving data streams. However, transferring the data to a centralized location is not feasible for privacy-preserving data analytics, as highlighted earlier. Aiming to generate a global model that can be deployed in new devices and represent the user's general behavior is of most interest. FL has been introduced for converging ML algorithms without considering the need for continuous updating. The challenge already occurring in FL with the communication overhead of transferring model updates and merged models over the network is increasing with the environment of constant learning. Even though FL is developed for non-independent and identically distributed (non-IID) data, deploying a generalized global model into devices for real-time prediction and actuating results in lower accuracy than having a local personalized model. However, customized models can overfit the data and produce in sudden changing environments poor prediction results compare to generalized models. The challenge lays in selecting in EDs the best and most accurate model to perform the analytics. Overcoming these issues, a regulation between global generalized and 
a 
local personal model is critical to implement to guarantee qualitative prediction at the device. These regulation needs to be efficient in terms of computation and energy as EDs are constrained and limited with power. Our research aims to provide an adaptive model weighting and time-optimized model selection at the edge using low computational complex algorithms for qualitative prediction results of the performed analytics. Privacy-efficient learning using a personalized FL methodology in combination with the traditional FL will enable the adaptation to distribution shifts while providing qualitative prediction results of contextual data streams in privacy-preserving environments.

\subsection{Use-Cases and Applications}
The importance of personalized and adapted learning in ED using privacy-preserving data will be highlighted 
by introducing a use-case 
and an application 
in which this is of high importance. The implementation of personal and generalized privacy-preserving learning is essential for 
an
application in which user and sensitive data is collected. This can be, for example, the automotive industry with the autonomous driving use-case. The car is equipped with several cameras placed inside and outside the vehicle. 
This generated data can contain situations 
where 
the driver and passenger are not happy sharing over the network to the central collection point for analysis and processing, e.g., a fight or argument. However, for companies, the data can benefit the quality of the user experience or even increase the safety. For 
instance,
the assistant can regulate the speed during the argument or activate safety functions for distance control. Not only are cameras 
one possible input data for autonomous driving applications but also sensor data of the vehicle parts or microphones placed inside the car using voice and emotion recognition. Combining the different data sources is 
the 
key for 
high 
quality and accurate applications. However, the transmission to adapt the model to the changing environment (e.g., a phone call that upset's the driver) is not feasible in real-time and the amount of data is costly even with the possibilities 
that 5G offers. Moreover, the vehicle is powered by either fuel or battery and therefore complex and computational 
intensive algorithms should be carefully considered. Personalized ML models that provide a unique driving experience are essential for the customer but with a change occurring in the environment (e.g., car-sharing with multiple drivers, weather change, personal emotion) that the personalized model never experienced, the accuracy of the predictive result decreases and the quality is not guaranteed anymore. The generalized model over all drivers can increase, during concept-drifts, the quality and accuracy, as it is equipped with more situations and data and fits even towards unseen data well.  Therefore, using local, in-car analytics with a privacy-efficient model selection methodology is important to provide a quality-aware user application. 

\subsection{Contribution}
To the best of our knowledge, we are 
among 
the first to propose the use of local personalized edge-centric analytics model implementation combined with FL approaches for generalization and concept shift adaptation using an adaptive model weighting and a time-optimized model selection. We summarize our contributions to this paper as follows:  
\begin{itemize}
\item{We investigate the importance of individual local edge-centric analytics across EDs 
for real-time prediction in contextual streaming data applications;}
\item{
We
design a continuous model re-training for evolving streaming data in EDs
incorporating FL for global knowledge generation for high accuracy of real-time prediction;}
\item{
We provide 
a novel intelligent and adaptive model weighting of the personalized local model and FL model across EDs
to improve the quality, accuracy, and robustness for concept drifts or behavior changes;}
\item{
We present
a method of a time-optimized model selection between personalized local and federated generalized models over continuously evolving data streams using statistical analysis based on the principles of Optimal Stopping Theory;}
\item{
We perform a 
comprehensive models 
evaluation and comparative assessment against current FL and global data acquisition approaches found in the literature using real-data sets.}
\end{itemize}

\subsection{Organization}
The following paper is structured and organized in introducing the fundamentals of 
FL
and related work in the field of personalized FL. The paper continues in presenting the rationale and 
the considered 
problems with highlighting the architecture and deployment of the introduced methodologies. 
We
continue with the detailed 
description and definitions of the proposed adaptive model weighting and the time-optimized model selection methodologies. Based on this introduced methodology, a performance assessment using real data is presented. This paper finishes with concluding the evaluated results and presenting future research topics given the presented work.

\section{Related Work}

Data security and privacy over EC 
environments has been summarized by Zhang et al. \cite{zhang2018data} where 
open issues and 
relevant 
topics re studied.
The problem of transmitting raw data over the network has been engaged from a security perspective by applying different edge deployable encryption algorithms.
The aspect of efficient privacy-preserving techniques for resource-constrained devices has been highlighted as an essential issue for further research. Our work does not contribute to security techniques but rather to efficient privacy-preserving machine learning and analytics in quality-aware edge networks. The focus relies on enabling edge device learning and training to preserve users sensitive data from transmitting. The aim is to keep the data at the location where it has been generated. This local learning and predictive analytics inside EDs can be seen as 
a
possible solution towards privacy-preserving analytics over sensible user data in edge environments. 

As highlighted, FL has been introduced as a methodology that has privacy by design. It has engaged the ED to learn on its local data a global ML
model without revealing any data towards a central 
entity.
FL has been extensively studied 
in scenarios where numerous 
IoT devices 
are present
(e.g., 
smartphones) using supervised learning and Neuronal Networks (NN). FL aims to learn a global function placed 
at a central location by pushing the training towards IoT devices. The central entity 
aggregates the gradients of all locally trained models together into a new global function using FedAvg \cite{McMahan2016, BrendanMcMahanEiderMooreDanielRamageSethHampsonBlaiseAgueraAg2017}. This form of distributed learning provides the possibility to use the computational power of devices and their locally generated data, creating beneficial aspects, such as privacy, real-time actuation, and robustness. FL can perform on massively distributed, non-IID, and unbalanced data a fast convergence of the global trained model.  

In many applications, generating a global model that is generalizing the data is of utter importance. However, in many IoT systems, the generated data is non-IID. In this case, the non-IID provokes personalized and local models to mostly outperform the generalized global model. This effect has been shown by the authors of \cite{Harth2018b, anagnostopoulos2020edge} 
performing local edge-learning to increase the quality of predictive analytics. Additionally, the authors in \cite{Kairouz2019} argue that one of the open research questions is when to choose the global model, providing generalization, over the local, providing individuality, and vice versa. A focus on personalization and FL has only recently 
attracted the 
attention of 
the research community. 
For instance, 
the authors in \cite{Kulkarni2020} are summarizing
the combination of personalized processing and 
FL. The authors in \cite{Zantedeschi2019, Vanhaesebrouck2017, Bellet2018, wang2019federated} show the impact of personalization in FL environments and its significant improvement towards qualitative prediction results. The research in \cite{Zantedeschi2019, Vanhaesebrouck2017, Bellet2018, wang2019federated} designs a fully decentralized architecture in which no global server coordinates the communication or has knowledge about the model gradients or the overall generated global model. 
EDs collaborate to share information and 
improve their model through other devices. This idea of decentralized personalised FL assumes complete connectivity between the different devices, which can not be assumed or realized in most IoT applications. Moreover, their research lacks in dealing with evolving data and concept drifts. Other methods proposed for personalized quality and privacy-efficient learning using FL rely on a hierarchical system order with a central coordinator performing the model merging. One of these related works is presented by Liang et al. \cite{Liang2019}. The authors use representation learning over the data features at each individual learner (i.e., ED). Their work provides individuality and personalization through having NNs 
with a lower-level global model locally enriched by a personalized representation layer. Meta-learning for FL, as a possibility to overcome the issue of heterogeneous data, has been introduced by the authors in \cite{fallah2020personalized, jiang2019improving}. They show the improvement in accuracy using personalized models in non-IID environments. 
Hanzely et al. \cite{Hanzely2020} propose a FL optimization 
technique 
that merges locally trained models to a global model providing a specialized gradient update procedure. Moreover, the authors Deng et al. \cite{Deng2020} present a combination of a globally learned model and a locally fine-tuned model. Deng et al. \cite{Deng2020} introduced an adaptive parameter in controlling the relationship between a global FL and a local fine-tuned model. In the 
proposed strategy, each device stores three models, the global, the local, and the personalized model. All of them have to be updated in each round of 
the adopted 
communication
activities. Their adaptiveness 
is calculated by the difference between the local and global model with respect to their gradient divergence. As their research is the most related work
to ours, 
we depart from them to propose a less computational complex procedure to adapt the relation between global generalized and local personalized models. With our method, no prior setting of the 
adopted parameters 
is required. The adaption is performed on a discrepancy-aware process that considers rewards of historical accuracy so that the model and data determine the parameter itself. Further, we rely on the assumption that convergence is not guaranteed. Therefore, continuous efficient model selection and updating are essential 
being
not considered in any mentioned literature.

Besides the personalization aspect of FL, we aim to use FL on contextual data with evolving nature. Performing FL on developing data is still an open research question, and little to no research has been done there. The closest related work is from Chen et al. \cite{Chen2019} proposing an online asynchronous version of FL. Their research suggests a convergence strategy of locally updating the FL model using feature representation learning at the central location and balancing with coefficients the relationship of previous gradients and current. We depart from their research by incorporating the importance of personalization and generalization using online FL in devices and adaptive model selection. However, we rely on the significance that FL has to develop in the devices between updates from the central location, as real-life applications continuously generate data. Moreover, 
Chen et al. \cite{Chen2019} do not consider 
the
efficiency and resource-constrained environments. This aspect is considered in the work of Wang et al. \cite{Wang2018}. They show the importance of local distributed learning in resource-constrained edge networks. The implementation of FL in a resource-constrained environment is presented in 
Sattler et al. \cite{Sattler2020}. They contribute 
by providing a communication efficient compression technique. The most recent work of Li et al. \cite{li2020lotteryfl} considers the essential requirement of using efficient communication methods in resource-constrained environments and the importance of personalization in a combined implementation. Yet again, the previous two mentioned methods assume the global convergence of a model without concept drifts in their work. The later work of Li et al. \cite{li2020lotteryfl} only improves communication by compression, sending fewer layers.  

\section{Rational \& Problem Fundamentals}
FL can be seen as a distributed ML 
optimization problem in a network of EDs. The aim is to locally optimize an objective 
function $\mathcal{J}$ over distributed datasets. A network of $k \in \{1,\ldots,k,\ldots, K\}$ EDs is designed with each holding a local subset of data $D_k$ so that $D=\{D_1, \ldots,D_k,\ldots, D_K\}$. 
$D_k$ is generated through Sensing and Actuating Nodes (SANs) sensing continuously contextual data in 
the
form of a $d$-dimensional vector $\mathbf{x}_t$ with $\mathbf{x} \in \mathbb{R}^d$ and $t \in \mathbb{T} = \{1,\ldots,T\}$ with $T \in \mathbb{T}$. 
At time instance 
$t$ a new contextual vector $\mathbf{x}_t$ is received at the ED. In centralized learning, this data is forwarded to a 
CL
that minimizes the objective function $\mathcal{J}$ over the entire dataset $D$. The optimization of the loss function $\mathcal{L}(\mathbf{x},y;w)$ can be approximated by exploring the Empirical Risk Minimization (ERM) instead of the expected risk. Therefore, an approximation of $\mathcal{\hat{J}}$ over a training set $D$ of size $N$ using a set of models $f \in \mathcal{F}$ with represented parameters $\mathbf{w} \in \mathbb{R}^d$ can find the optimal solution for the ERM. This has been highlighted in the following equation:
\begin{equation}
   \arg \min_{w \in \mathcal{F}} \mathcal{J}(w) \approx \mathcal{\hat{J}}_N(w) = \frac{1}{N} \sum^{N}_{i=1} \mathcal{L}(x_i, y_i; w)
   \label{eq:ch5_objectivefunction}
\end{equation}

In FL, the optimization of 
$\mathcal{J}$ is performed by locally optimizing a objective function $\mathcal{J}_k$ in 
the $k$th
ED 
over 
$D_k$ with length $n_k$ so that $\sum^K_{k=1} n_K= N$. The CL is aggregating the local generated objective functions $\mathcal{J}_k$ so that:
\begin{equation}
   \mathcal{J}(w) = \sum^{K}_{k=1} \frac{n_k}{N} \mathcal{J}_k(w) = \sum^{K}_{k=1} \frac{n_k}{N} \frac{1}{n_k} \sum_{i \in D_k} \mathcal{L}(x_i,y_i,w)
   \label{eq:ch5_objectivefunction_ed}
\end{equation}

The loss function $\mathcal{L}(x_i,y_i,w)$ or local optimization function $\mathcal{J}_i(w)$ can be also noted as $j_i(w)$. One algorithm to solve the problem
depicted by 
Equation (\ref{eq:ch5_objectivefunction_ed}) is using the previous introduced Stochastic Gradient Descent (SGD). 
At 
each iteration $t$, the $k$th ED 
aims to converge towards the minimum loss function $\mathcal{L}$ over its local data $D_k$ 
given a new datapoint $\mathbf{x}_t$. This is by adapting the model parameters $\mathbf{w}_k^{t}$ with some factor $\eta$, called learning rate, and the gradient to the previous model parameters $\nabla j_{k}(\mathbf{w}_k)$. 
The $k$th ED
is performing the following processing at 
$t$:
\begin{equation}
    \mathbf{w}_k^{t+1} \leftarrow \mathbf{w}_k^t - \eta \nabla j_k(\mathbf{w}_k^t)
\label{eq:ch5_SGDw}
\end{equation}

After multiple iterations, the updated model parameters of the $k$th ED
are sent to the CL. The introduced FedAVG algorithm \cite{McMahan2016} aggregates the received model parameters inside the central coordinator towards the new generalized model. This averaging is summarized 
by the following Equation: 
\begin{equation}
    \mathbf{w}^{t+1} = \sum_{k=1}^{K} \frac{n_k}{N}\mathbf{w}_k^{t+1}
\label{eq:ch5_fedavg}
\end{equation}

After merging the local gradients of the $k$th ED
at the CL, the final federated model is distributed to the EDs and used locally for inference. 
The $k$th ED
is selected at a random time epoch $s$ with $s=\{1,\ldots,s,\ldots,S\}$ to update the distributed federated model $f_{FL}$ with its local data stored in a Sliding Window (SLW) $W_k^t$ and send the updated model parameters $\mathbf{w}_k$ back to the CL. 

\begin{figure}[h]
\includegraphics[width=0.8\columnwidth]{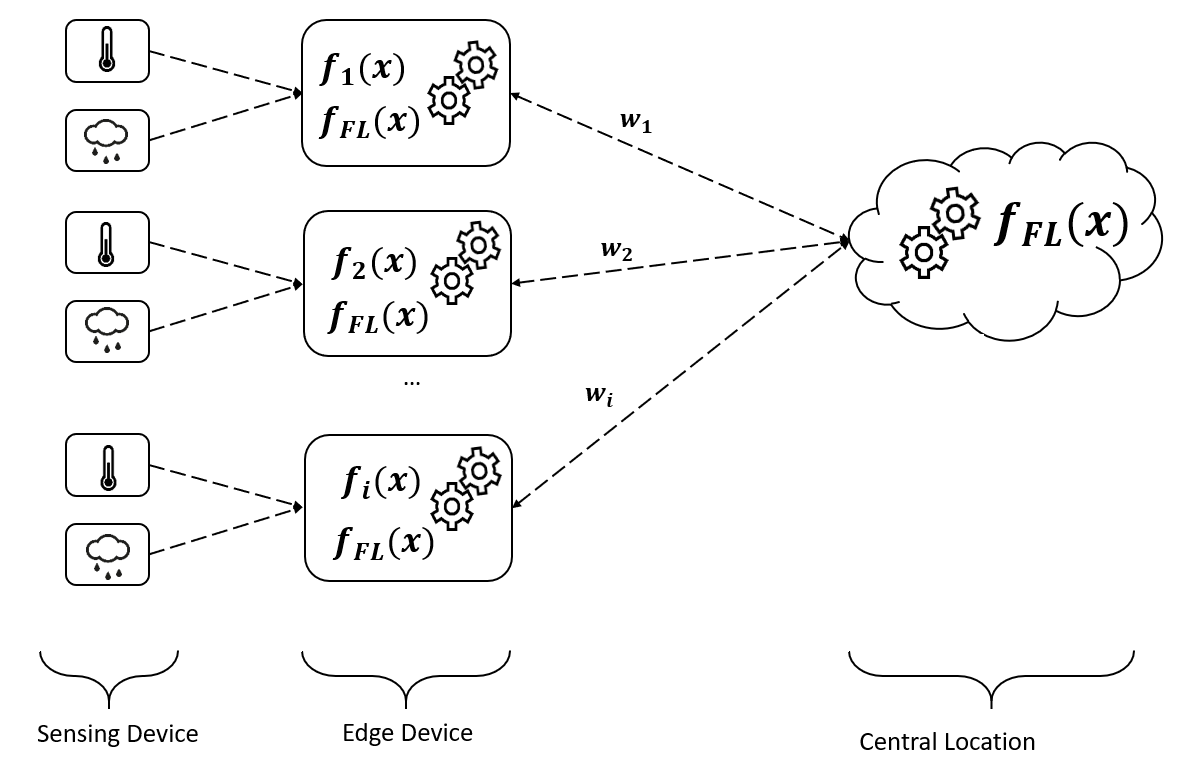}
\caption{Generating global knowledge with FL in Edge Devices and distribution of locally trained models to the CL.}
\label{fig:architecture}
\end{figure}

Highlighted in the State-of-the-Art section, the basic implementation of FL introduces multiple problems. The first issue is the adaptation 
of the generalized model to constantly changing environments, in which the assumption of converging to a global minimization of the objective function $\mathcal{J}$ does not hold. The second problem arises as the ED always overwrites the locally adapted model $\hat{f}_{FL}$ with a newly received generalized model $f_{FL}$. The issues of generalized models of non-IID data have been highlighted in the previous section, showing the importance of personalization for qualitative analytics at CLs.
Therefore, this paper 
introduces the dual model deployment inside 
EDs.
The $k$th ED 
implements an evolving personalized model $f_k$ in parallel to the generalized federated model $f_{FL}$. 
An overview of the constructed network architecture can be seen in \autoref{fig:architecture}.  

The major challenge inside each ED 
with the deployed dual model methodology is 
to
choose 
the correct model for the prediction $\hat{y}$ with either $\hat{y}=f_{FL}(\mathbf{x})$ or $\hat{y}=f_{k}(\mathbf{x})$. 
EDs are 
limited with their resources, which does not allow complex algorithms to be implemented for decision making locally. Therefore, in the following sections, lightweight methods for balancing the personalized and generalized federated model inside each ED for qualitative analytical results are presented. Moreover, current deployed and tested FL methodologies consider a supervised learning environment. In many applications, e.g. autonomous cars or smart homes, the label to check the performance of the supervised learning algorithm is not provided. Therefore, systems and algorithms using continuous data prediction or forecasting with a FL deployment are of high interest in industrial implementation.

\section{Personalized Privacy-Efficient Learning at the Edge}

Highlighted in the 
previous 
sections, 
personalized FL models inside the ED has gained tremendous interest in the last years to overcome the problems of heterogeneous non-IID data. The open issues of continuously changing data and concept drift adaptation 
have been tried to overcome by using a local evolving personalized model inside the ED. The authors in \cite{Harth2018b, anagnostopoulos2020edge} highlighted, the importance of local models inside EDs with selecting from an ensemble of models at the CL has been proven to generate qualitative and efficient analytics for query-driven predictions. Using local personalized models inside the ED and performing ensemble pruning strategies, is due to restrictions of sharing user-specific models or meta-data to the CL impossible. The introduced general method of FL 
shows the process of generating a model on the CL by training on distributed datasets without accessing the raw data on each ED, so having the privacy of the data by design. However, the drawback of generalized models implemented in the FL setup can be overcome by local build models. This results in the problem that by only having a local personal model at the ED, the adaptation towards unseen data induced by concept shift will generate inaccurate and wrong predictions. Therefore the following work focuses on using the combined power of generalization using the FL model and the advantage of the personalization model inside each ED. This should provide qualitative prediction over changing environments in edge networks while being resource-efficient and reducing the communication and complexity. The basic implementation of FL has a complexity inside the EDs with $O(n_k(d+1))$ and at the CL with $O(k+d+1)$.

Having a parallel model deployment of the federated model and a personalized model inside the EDs requires first defining how and when to update both models and second which model to choose from for qualitative predictions at the ED. As mentioned earlier at epoch $s$, 
the $k$th ED
is selected to receive the current generalized 
FL
model $f_{FL}$ from the CL. Then, the ED performs over its locally stored data in a SLW $\mathcal{W}_k$ the model gradient updates using Equation (\ref{eq:ch5_SGDw}). These updates are forwarded to the CL and used into the new model $f_{FL}$ combining all selected $k$ EDs model parameter updates with FedAVG, as shown in Equation (\ref{eq:ch5_fedavg}). 
All EDs receive the updated federated model after the merging and can use this as a model to perform the predictive analytics locally. 

The personalized model $f_k$ is set to be at $t=0$ the centrally received federated model $f_{FL}$. The choice of using a pre-build model as starting point lies in the theories of transfer learning, \cite{torrey2010transfer, pan2009survey} which highlights that using a baseline model for retraining is generating an earlier convergence and higher accuracy than starting from scratch.  In the majority of IoT or edge environments, SANs continuously measures the surrounding. The ED receives this data from the SANs at each time $t$ with $t \in \mathbb{T} = \{1,\ldots,t,\ldots,T\}$ and $T \in \mathbb{T}$. Each time $t$ a $d$-dimensional data vector $\mathbf{x}_t$ is collected and stored in the local window storage $\mathcal{W}_k$ of size $M$. Only at the selected epoch $s$ this data is used to update the newly received model $f_{FL}$. If the ED needs to perform a prediction, it is using the previous cached model. The local model $f_k$ is in compare updated and retrained each time $t$ at the ED using the local SGD process. The retraining methodology of \cite{Harth2018b} is additionally possible to use for energy-efficient retraining inside the ED but is not further considered in this research paper. 

The aim is to present privacy-efficient federated learning methodologies improving the current work by deploying personalized aspects and balancing strategies inside the continuous changing environment of EDs. Therefore, in the following, the two strategies are illustrated contributing to personalized  privacy-efficient learning in edge networks based on the fundamentals of FL. 

\textbf{Evolving Federated Model (EFM)}:
The first method of incorporating personalization into a privacy-preserving edge environment is advancing on the basic implementation of FL. The EFM introduces a communication-efficient strategy that discards the updating communication from the CL to the ED after generating the new generalized model $f_{FL}$. The ED $k$ received at the selected epoch $s$ the model $f_{FL}$ and updates the model parameters using Equation (\ref{eq:ch5_SGDw}), then it communicates these updated model parameters to the CL. Instead of receiving the new merged model $f_{FL}$, the ED keeps the updated model as the $f_k$ and continuously evolves the model each time $t$ a new measurement $\mathbf{x}_t$ is collected. If 
the $k$th ED
is selected at another epoch $s$, 
$f_k$ is replaced with the new $f_{FL}$. In this strategy, the communication is reduced, and the ED can adapt to the evolving data and concept drifts by continuously learning and retraining the generalized model to its personalized environment. 

\textbf{Local Federated Model (LFM)}:
The second introduced privacy-efficient FL based methodology is limiting the communication further between 
EDs and the CL. The concept is to initialize at time $t=s=0$, the local model $f_k$ with the received federated model, so that $f_k=f_{FL}$ inside each ED. 
The ED is, then, continuously updating the model using SGD, see Equation (\ref{eq:ch5_SGDw}). Instead of the CL requesting each epoch $s$ the update of the distributed $f_{FL}$, the ED is regularly sending the local model $f_k$ to the CL. Inside the CL, the model parameters of each $f_k$ are merged into a global model $f_{FL}$. This can be distributed if the divergence is too large or if new devices join the network. 
This type of personalized local FL has been partly introduced in the related work of \cite{Liang2019}. The authors train local representatives of higher network layers and merging them centrally with the original deeper layers, thus,
no distribution of merged models is required anymore.

\section{Adaptive Model Weighting}

The introduced personalization adaption of FL in the above section only presents the use of a single model inside the ED that is adapting and continuously evolving. In the following, we propose an adaptive model selection methodology performed inside each ED. This 
methodology uses a reward system on historical accuracy to weight the relation between local and federated models. 
The argument and necessity of having an adaptive model selection on the ED is to the unknown IID or non-IID data relationship observed in 
many applications.
When having an adaptive model selection, a high accuracy independent on the relationship of the distributions can be provided for predictive analytic tasks. Further, considering the frequent appearance of concept and distribution shifts for contextual and evolving data, it is important to have besides a personalized model, another model that represents a generalization of 
data, 
especially when focusing on the quality of the prediction. 

The following method of privacy-efficient personalized local learning introduces the combination of keeping two models in parallel inside the ED. As LFM and EFM only store one model and continuously retraining it, the Adaptive Selection Method (ASM) methodology supports the generalized model $f_{FL}$ and the local model $f_k$ inside each ED $k$. The importance of keeping both models inside the ED is mainly to overcome the issue of losing generalization and the adaption to unseen data by only deploying personal models. Moreover, the previously unknown relationship between the devices in evolving systems can be changing from an IID to a non-IID data relationship. This cannot be identified at the start of the application deployment and can cause a better accuracy to the generalized model or personalized model depending on the connection. Using a balancing mechanism of the generalized and personalized model inside each ED, we can provide the benefits of both data relations towards qualitative prediction results. The first introduced this strategy are the authors in \cite{Deng2020}. However, their balancing of both models relies purely on the gradient difference between these two models 
not including previous predictions or 
considering the adaptation 
on changing environments. As this adaptation is of high importance to most time series analytics in IoT and edge environments, the ASM contributes to the research gap of using a reward system to weigh the local and federated model into a combined prediction $\hat{y}$. The final prediction $\hat{y}$ is calculated locally in each ED 
by using the local personalized function $f_k$ and the FL-function $f_{FL}$ with an adaptive balancing weight $\alpha$. 

The adaptive weight is calculated through multiple steps. First, whenever the ED receives from the SAN a new contextual vector $\mathbf{x}_t$ at time $t$, the prediction error $\epsilon_k$ for the local, personalized model $f_k$ with respect to the actual prediction $y_t$ and the prediction error $\epsilon_{FL}$ for the federated model $f_{FL}$ are 
calculated as shown in Equations (\ref{eq:ch5_e_local}) and (\ref{eq:ch5_e_federated}), respectively. 
\begin{equation}
\epsilon_L = |y_t-f_k(\mathbf{x_t})|,
\label{eq:ch5_e_local}
 \end{equation}
\begin{equation} 
\epsilon_{FL}=|y_t-f_{FL}(\mathbf{x_t})|.
\label{eq:ch5_e_federated}
 \end{equation}

Given these two errors, 
inside each ED, 
it is possible to generate a reward value $\theta$ that is used as the factor for balancing the two models $f_k$ and $f_{FL}$. Hence, in each ED 
at 
$t$, based on $\epsilon_{FL}$ and $\epsilon_L$, the reward $\theta$ is set to $0$ if the local model is performing better than the generalized, and $\theta=1$ if the generalized model $f_{FL}$ performs better predictions. This reward setting is depicted by 
the following equation:
 \begin{equation}
     \theta_t =
    \begin{cases}
    0, & \quad \epsilon_{FL} > \epsilon_L  \\
    1 & \quad \epsilon_{FL} \leq \epsilon_L. 
    \end{cases} 
\label{eq:ch5_theta}
 \end{equation}

As shown in Equation (\ref{eq:ch5_theta}), a positive reward is given to the federated model $f_{FL}$ if the absolute error $\epsilon_{FL}$ is smaller than the absolute error of the local model $f_k$ $\epsilon_L$. As the reward system deployed inside the ED should not only incorporate the current performance of $f_k$ and $f_{FL}$ but, also, the historical performance, the reward-values $\theta_{t}$ 
are stored in a SLW $\mathcal{O}_{k}$ with size $U$ for the last $t-U$ time instances. The SLW is defined in Equation (\ref{eq:ch5_slw_theta}) and contains zeros and ones.  

\begin{equation}
\mathcal{O}_{k}^t = \{\theta_{t-U}, \ldots, \theta_{t} \}.
\label{eq:ch5_slw_theta}
 \end{equation}
 
The deployment of a SLW is chosen 
towards 
the adoption of 
lightweight methods for handling continuous evolving data and adaptation to changing environments able to act immediately on concept drifts \cite{Bifet2007LearningWindowing, dietterich2002machine}. The SLW $\mathcal{O}_{k}$ consists of the most recent $U$ rewards of $\theta \in \{0,1\}$. At each time $t$, the SLW of rewards is used as a reference to generate a ratio that represents the performance of both models over the time horizon $t-U$. 
This ratio is defining the adaptive weighting value $\alpha$. The computation of $\alpha$ is given in Equation (\ref{eq:ch5_alpha}) and represents the historical performance of both models inside the ED.

\begin{equation}
    \alpha = \frac{1}{U} \sum _{i=1}^{U}\theta_i.
\label{eq:ch5_alpha}
\end{equation}

Each time the ED is performing a prediction, 
$f_k$ and $f_{FL}$ are weighted to the final prediction $\hat{y}$. The adaptive value $\alpha$ is used to combine the federated model $f_{FL}$ and the local personalized model $f_k$ towards $\hat{y}$ using exponential smoothing \cite{Durbin2002AAnalysis}. The balancing of these two models is defined as shown in Equation (\ref{eq:ch5_asm}):
\begin{equation}
    f_{ASM} = \alpha f_{FL} + (1-\alpha) f_k
\label{eq:ch5_asm}
\end{equation}

 \begin{algorithm}[h!]
\caption{Adaptive Smoothing Model}
 \begin{algorithmic}[1]
 \renewcommand{\algorithmicrequire}{\textbf{Central Location:}}
 \REQUIRE 
 \STATE initialize $\mathbf{w}_0$
 \FOR{each round $s=1,2,...$}
 \STATE $i =$ random subset of $K$
 \FOR{each ED $i$ \textbf{in parallel}}
 \STATE $\mathbf{w}_i^{s+1} \leftarrow ed(i,\mathbf{w}_s)$
 \STATE $\mathbf{w}_i^{s+1} \leftarrow \sum_{i=1}^{K} \frac{n_k}{N} \mathbf{w}^{s+1}_i$
 \ENDFOR
 \ENDFOR
 \item[]
 \renewcommand{\algorithmicensure}{\textbf{Edge Device:}}
 \ENSURE  \textit{ //Run on each Edge Device $k$}
 \STATE $f_{FL} \leftarrow$ (received from cl at $t=0$)
 \FOR{each round $t=1,2,...$}
 \STATE contextual vector $\mathbf{x}_t$ is received
  \STATE $f_k \leftarrow$ updated via Equation (\ref{eq:ch5_SGDw})
  \STATE $\epsilon_L \leftarrow |y_t-f_k^t(\mathbf{x}_t)|$
  \STATE $\epsilon_{FL} \leftarrow |y_t-f_{FL}^t(\mathbf{x}_t)|$
  \STATE $\theta \leftarrow$ calculated as in Equation (\ref{eq:ch5_theta})
  \STATE $\mathcal{O}_t \leftarrow$  $\theta$
  \STATE $\alpha \leftarrow$ calculated as in Equation (\ref{eq:ch5_alpha})
  \IF{$\hat{y}$ is needed}
  \STATE $f_{ASM} \leftarrow$ calculated as in Equation (\ref{eq:ch5_asm}) 
  \ENDIF
  \IF{$s = t$ and ED $i = k$}
  \STATE $f_{FL} \leftarrow$ updated as in Equation (\ref{eq:ch5_SGDw}) or use $\nabla f_k$
  \STATE return updated $\mathbf{w}$ to cl
  \ENDIF
 \ENDFOR
\end{algorithmic} 
\label{algorithm:ch5_ASM}
\end{algorithm}

The value range of $\alpha$ is 
in the interval $[0,1]$. 
Setting $\alpha \longrightarrow 1$ indicates that more influence towards the FL model predictions coming from $f_{FL}$, whereas a value of $\alpha \longrightarrow 0$ places more importance on the locally generated model $f_k$ predictions. The proposed adaptive model selection 
is highlighted by 
the Algorithm \ref{algorithm:ch5_ASM}, showing the steps of calculating the reward $\theta$ and ratio $\alpha$, which are 
used to combine the generalized model $f_{FL}$ and local personalized model $f_k$ towards a weighted and more accurate analytical model inside each ED. 

\section{Time-Optimized Model Selection}

\subsection{Fundamentals of Optimal Stopping Theory}
The fundamentals of Optimal Stopping Theory (OST) \cite{shiryaev2007optimal, Robbins_OST} lay in choosing a time to take a action that maximizes an expected return. The rule of stopping at this time instance is defined by having a sequence of random variables $Z_1,Z_2, \ldots$ and a sequence of rewards that depend on the observed value of $Z$ until time $t$, so that the sequence of return functions can be defined as $(Y_t(Z_1,\ldots,Z_t))_{t > 1}$. The application is observing 
the sequence of $Y_t$ and decides to either stop or continue. In our method, the application decides to switch the model between the federated and the local and vice versa. Our aim is to maximize the expected return or reward of the function $Y_t$ when we decide to switch the models.

\subsection{Privacy-Efficient Model Selection}

The 
already introduced idea of rewarding historical performance values towards balancing the local model $f_k$ and the generalized federated model $f_{FL}$ by weighting the two predictions towards the final prediction $\hat{y}$ has been shown in ASM. Relying on the concept of incorporating historical decisions with respect to the performance, the Time-Optimized Selection Method (TOSM) introduces the concept of OST. The design of TOSM is to overcome the problem of selecting between the two models $f_k$ and $f_{FL}$ inside the ED. Instead of weighting the predictions, just one model is chosen for the prediction $\hat{y}$ by identifying the \textit{optimal} model at this time $t$. 

The main concept of TOSM is to calculate inside each ED 
at time $t$ for each local model $f_k$ and federated model $f_{FL}$ the prediction errors $\epsilon_L$ and $\epsilon_{FL}$ as defined in Equations (\ref{eq:ch5_e_local}) and (\ref{eq:ch5_e_federated}), respectively. At time $t=s=0$, the chosen model inside the ED is the $f_{FL}$. Given the two prediction errors $\epsilon_L$ and $\epsilon_{FL}$, the idea is to decide when 
is 
the best time $t^*$ 
to switch from the federated model $f_{FL}$ to the local model $f_L$. 
The authors in \cite{Harth2018d} adopt OST through the reconstruction error difference for finding the optimal time to forward the raw data from the SAN to the ED. For TOSM, the fundamental idea 
is used and performed on the prediction error $\epsilon_{FL}$ and $\epsilon_{L}$ as performance value. Each time $t$ the two prediction errors of $f_k$ and $f_{FL}$ are compared and rewarded as a binary value $Z$, i.e., 
the binary variable $Z$ is introduced in Equation (\ref{eq:ch5_Z}).
\begin{equation}
Z_t =
\left\{
	\begin{array}{ll}
		\theta=0 &  \mbox{if } \epsilon_{L} > \epsilon_{FL}, \\
			\theta= 1 & \mbox{if } \epsilon_{L} \leq \epsilon_{FL}.
	\end{array}
\right.
\label{eq:ch5_Z}
\end{equation}

$Z_t$ is defined as 0 if the local model $f_k$ is generating a higher prediction error than the federated model. If the local model is, however, generating a lower prediction error $\epsilon_{L}$ in comparison to the federated model $f_{FL}$, the value of $Z_t$ is set to be 1. Assume that the ED is deciding instantly based on the given comparison between $\epsilon_{FL}$ and $\epsilon_{L}$ 
to switch the model. In that case, no historical behavior is incorporated towards this decision. As previous behavior and prediction quality are of importance to most applications, the values of the cumulative sum of comparison, including the history divergence of these two models, are used as switching decision. The instruct history of rewarded prediction performance comparison is then calculated by the cumulative sum of $Z_t$ defined as $R_t$ and introduced as:

\begin{equation}
R_t = \sum_{i=0}^t Z_i.
    \label{eq:ch5_sum_Z}
\end{equation}

In \cite{Harth2018d}, 
a
proof has been given that by defining the reward function $Y_t$, 
thus, 
it is possible to construct the optimal delay-tolerant level between SAN and ED to forward data. In the previous section, the accumulated sum of reconstruction errors has been used as the reward function $Y_t$. Adopting this concept, the accumulated sum of $Z_t$ defined as $R_t$ is used to find the optimal time to decide when to switch from the federated model $f_{FL}$ to the local model $f_k$. Therefore the reward function for switching $f_{FL}$ to $f_k$ is defined in Equation (\ref{eq:ch5_reward_ost_Z}) using 
$R_t$. 
$\beta \in (0,1)$ indicates the (delay) tolerance level. 
If $\beta \rightarrow 1$ the tolerance is increased. 

\begin{equation}
Y_t = \beta^t R_t = \beta^t \sum_{i=0}^t Z_i.
    \label{eq:ch5_reward_ost_Z}
\end{equation}

To
find
the optimal stopping time $t^*$, the reward function $Y_t$ depicted by 
Equation (\ref{eq:ch5_reward_ost_Z}) is used by maximizing the expectation of $Y_t$ with $\mathbb{E}[Y_t]$ having a fixed tolerance of $\beta$. Formally, this can be described as finding the supremum of the expectation of $Y_t$: 
\begin{eqnarray}
\sup_{t \geq 0}\mathbb{E}[Y_{t}].
\label{eq:ch5_sup_expectation_y}
\end{eqnarray}

Proofing that the optimal time $t^*$ exists 
can be shown using the fundamentals of OST. Two conditions need to be satisfied: (C1) $\lim \sup _{t} Y_{t} \leq Y_{\infty} = 0$ is surely true and (C2) $\mathbb{E}[\sup_{t} Y_{t}] < \infty$.
C1 implies that with the elapse of time ($t \to \infty$), the reward should go to zero, i.e., $Y_{\infty} = 0$. Since no change of the model over an indefinite horizon is useless due to placement in constantly changing environments, 
$Y_{\infty} = 0$ represents the reward of an endless non-delivery phase. The supremum limit of $Y_{t}$ is notated by $\lim \sup_{t} Y_{t}$, i.e., the limit of $\sup_{t} Y_{t}$ as $t \to \infty$ or $\lim_{t \to \infty}(\sup\{Y_{j} : j \geq t\})$. As $Z_{t}$ is non-negative and using the strong law of numbers $(\frac{1}{t}\sum_{j=1}^{t}Z_{j}) \to \mathbb{E}[Z]$ it is possible to derive: 
\begin{equation}
Y_{t} = t\beta^{t}(R_{t}/t) = t\beta^{t}(1/t)\sum_{j=1}^{t}Z_{j} \sim t \beta^{t} \mathbb{E}[Z] \overset{\mbox{a.s.}}{\to} 0.
\label{eq:ch3_lim_yt}
\end{equation}

This results to $\lim_{t \to \infty} \sup_{t} Y_{t} = 0$. As $Y_{\infty} = 0$ is by definition true, it is possible to declare C1 as satisfied. C2 implies that the expected reward under any policy is finite. Therefore, C2 can be shown as: 
\begin{equation}
\sup_{t} Y_{t} = \sup_{t}\beta^{t}\sum_{j=1}^{t} Z_{j} \leq \sup_{t}\sum_{j=1}^{t}\beta^{j}Z_{j} \leq \sum_{j=1}^{\infty}\beta^{j}Z_{j}. 
\label{eq:ch3_sup_yt_proof}
\end{equation}
This results into satisfying C2 with, 
\begin{equation}
\mathbb{E}[\sup_{t}Y_{t}] \leq \sum_{j=1}^{\infty}\beta^{j}\mathbb{E}[Z] = \mathbb{E}[Z]\frac{\beta}{1-\beta} < \infty.
\label{eq:ch3_expected_sup_yt}
\end{equation}
As both conditions are satisfied and proven, it can be shown that the optimal time $t^*$ for forwarding the measurements in Equation (\ref{eq:ch5_sup_expectation_y}) exists.

Proofing of the existence of the optimal time $t^*$ desires to find that optimal time inside the ED which enables it to decide on switching the models by maximizing the trade-off between their accuracy.
Since $Y_{t}$ is non-negative, the Equation (\ref{eq:ch5_sup_expectation_y}) is monotone \cite{Robbins_OST} so that the optimal time $t^{*}$ is obtained by the one-stage look-ahead optimal rule (1-sla): 
\begin{equation}
t^{*} = \inf\{t \geq 1 | Y_{t} \geq \mathbb{E}[Y_{t+1}]\}.
\label{eq:ch3_opti_time_1}
\end{equation}
The adoption of 1-sla is optimal since $\sup_{t} Y_{t}$ has a finite expectation (equal to $\mathbb{E}[Z]\frac{\beta}{1-\beta}$) and $\lim \sup_{t} Y_{t} = 0$, as proved in Equation (\ref{eq:ch3_lim_yt}).
Consequently, $t^{*}$ is estimated through the principle of optimality. 
Presume $R_{t} = r$ when a ED decides that it is optimal to switch the model. 
Then, the current reward of $\beta^{t}r$ is at least as large as any 
expected $\mathbb{E}[(\frac{\beta}{1-\beta})^{t+\tau}(r+R_{\tau})]$,
which means that: $s(1-\mathbb{E}[(\frac{\beta}{1-\beta})^{\tau}]) \geq \mathbb{E}[(\frac{\beta}{1-\beta})^{\tau}R_{\tau}]$ for all times $\tau$. This must hold true for all $r' \geq r$, so that 
the optimal time $t^{*}$ for some $r_{0}$ must be of the form $t^{*} = \inf\{t \geq 1 | R_{t} \geq r_{0}\}$. 
Especially when the ED switches the first time $t$ for which $R_{t} \geq r_{0}$, then the tolerance for forwarding $r_{0}$ 
must be the same as the tolerance for continuing using the 1-sla, therefore the sum of tolerances is positive. That is, $r_{0}$ must satisfy the equation 
\begin{equation}
s_{0} = \mathbb{E}[(\frac{\beta}{1-\beta})^{\tau}(r_{0} + R_{\tau})], 
\label{eq:ch3_s0}
\end{equation}
with $\tau=\inf\{t \geq 1| R_{\tau} > 0\}$. Since $Y$ is non-negative, it is possible to obtain $\tau \equiv 1$ and $S_{\tau} \equiv Y$ \cite{Robbins_OST} and, then, replacing with $s_{0} = \frac{\beta}{1-\beta}\mathbb{E}[Y]$. This will finally result in the definition of the optimal time $t^*$ for switching from the federated model $f_{FL}$ to the local model $f_k$ 
defined as:
\begin{equation}
t^{*} =  \inf\{t \geq 1 | \sum_{i=1}^{t}Z_{i} \geq \frac{\beta}{1-\beta}\mathbb{E}[Z]\}.
\label{eq:ch5_optimal_time_switch}
\end{equation}

The expectation $\mathbb{E}[Z]$ can be calculated upon 
$Z_t$ 
formulated 
in Equation (\ref{eq:ch5_Z}) and 
the summation of the expectation $\mathbb{E}[Z|\epsilon_{L}>\epsilon_{FL}]$ with 
$\mathbb{E}[Z|\epsilon_{L} \leq \epsilon_{FL}]$ presented in the following equation:

\begin{eqnarray}
\!\!\!\!\!\!\! \mathbb{E}[Z] & \!\! = & \!\!\!\!\! \mathbb{E}[Z|\epsilon_{L}>\epsilon_{FL}]P(\epsilon_{L} > \epsilon_{FL}) + \\ \nonumber
& &\mathbb{E}[Z|\epsilon_{L} \leq \epsilon_{FL}]P(\epsilon_{L} \leq \epsilon_{FL}) \\  \nonumber
\label{eq:ch5_expectation_Z_1}
\end{eqnarray}

As the expectation of $Z$ is given 
by
Equation (\ref{eq:ch5_Z}) set to $\mathbb{E}[Z|\epsilon_{L}>\epsilon_{FL}] =0$ and  $\mathbb{E}[Z|\epsilon_{L} \leq \epsilon_{FL}] = 1$, 
the overall expectation of $\mathbb{E}[Z]$ can be defined as:  
\begin{eqnarray}
\!\!\!\!\!\!\! \mathbb{E}[Z] & \!\! = & \!\!\!\!\! P(\epsilon_{L} \leq \epsilon_{FL})= P(\epsilon_{FL} \geq \epsilon_{L}) \\  \nonumber
& = & 1- P( \epsilon_{L}<\epsilon_{FL})= 1- F_{\epsilon_{L}}(\epsilon_{FL})\\  \nonumber
\label{eq:ch5_expectation_Z_2}
\end{eqnarray}

During a training period of the application, the Probability Density Functions 
(PDFs) 
of $\epsilon_{FL}$ and $\epsilon_{L}$ can be obtained, which leads to generate the optimal time $t^*$ for switching the federated model $f_{FL}$ to the local model $f_k$ to:
\begin{equation}
t^{*} =  \inf\{t \geq 1 | \sum_{i=1}^{t}Z_{i} \geq \frac{\beta}{1-\beta}(1-F_{\epsilon_{L}}(\epsilon_{FL})\}.
\label{eq:ch5_optimal_time_switch2}
\end{equation}

The introduced methodology 
is further considering the reverse switching from the local model $f_k$ to the generalized model $f_{FL}$. Therefore, once the model is switched from the federated model to the local one, 
$R$ is set to 0. Then, the concept of finding the optimal time $t^*$ to switch back to the federated model with using the same concept as previous is starting inside the ED.
The optimal time to switch from the local model $f_k$ to the generalized model $f_{FL}$ is made through the variable $Q$ through 
which the prediction errors $\epsilon$ for both models are monitored. Similar to 
$Z_t$, $Q_t$ is calculated as:
\begin{equation}
Q =
\left\{
	\begin{array}{ll}
			\theta= 0 &  \mbox{if } \epsilon_{FL} > \epsilon_{L}, \\
			\theta=1 & \mbox{if } \epsilon_{FL} \leq \epsilon_{L}.
	\end{array}
\right.
\label{eq:ch5_Q}
\end{equation}

From Equation (\ref{eq:ch5_Q}), the expectation of $Q$ using the ideas of Equation (\ref{eq:ch5_expectation_Z_1}) and (\ref{eq:ch5_expectation_Z_2}) can be delivered by: 

\begin{eqnarray}
\!\!\!\!\!\!\! \mathbb{E}[Q] & \!\! = & \!\!\!\!\!  \mathbb{E}[Q|\epsilon_{FL}>\epsilon_{L}]P(\epsilon_{FL} > \epsilon_{L}) + \\ \nonumber
& & \mathbb{E}[Q|\epsilon_{FL} \leq \epsilon_{L}]P(\epsilon_{FL} \leq \epsilon_{L}) \\\nonumber
& = & P(\epsilon_{FL} \leq \epsilon_{L})= P(\epsilon_{L} \geq \epsilon_{FL}) \\  \nonumber
& = & 1- P( \epsilon_{FL}<\epsilon_{L})= 1- F_{\epsilon_{FL}}(\epsilon_{L})\\  \nonumber
\label{eq:ch5_expectation_Q}
\end{eqnarray}

From the given expectation of $Q$ in Equation (\ref{eq:ch5_expectation_Q}), it is possible to derive the optimal time $t^*$ to switch from the local $f_k$ back to the generalized $f_{FL}$. Only the PDF of the error values $\epsilon$ of the models $f_k$ and $f_{FL}$ is needed and can be generated through a training period of the application inside each ED.
In Equation (\ref{eq:ch5_optimal_time_switch_lfl}), the optimal time $t^*$ for the switch is defined. Once the model is switched back to the generalized $f_{FL}$, the summation of values of $Q$ is reset, and the $Z$ accumulation is starting. This method is performed until a new federated model $f_{FL}$ is sent from the CL, 
replacing the old one. This provokes a reset and a starting of using $Z$ with the current chosen model $f_{FL}$.
\begin{equation}
t^{*} =  \inf\{t \geq 1 | \sum_{i=1}^{t}Q_{i} \geq \frac{\beta}{1-\beta}(1-F_{\epsilon_{FL}}(\epsilon_{L})\}.
\label{eq:ch5_optimal_time_switch_lfl}
\end{equation}

\section{Performance Evaluation}
\label{sec:experiments}
\subsection{Experimental Setup}
For assessing the performance of our proposed methodology adaptive model selection, we performed all experiments on a dataset (DS) that contains 415 weather stations around the United Kingdom (UK) measuring contextual data of the surrounding environment. This data has been collected over the time horizon of December 2017 till March 2018 using the API of Wunderground \cite{WeatherUnderground}. Each weather station represents the $k$th 
ED 
with $k \in K = \{1,\ldots,k\ldots,K\}$ so that $K=415$. Each time $t$, 
EDs received a $d$-dimensional input data vector $\mathbf{x}_t$. The DS provides a $9$-dimensional data in 
the
form of $(\mathbf{x},y)$ 
including temperature, dew point, humidity, wind-speed, wind-gust, wind direction, pressure, windchill, and precipitation.  
The value for $y$ is set with the DS's measurement of temperature, while the remaining measurements are used for $\mathbf{x}$. The function $y = f(\mathbf{x})$ is a multivariate linear regression with $f(\mathbf{x}) = \mathbf{w}^T\mathbf{x}; \mathbf{w} \in \mathbb{R}^d$ resulting in minimizing the objective function of Equation (\ref{eq:ch5_objectivefunction}) to:
\begin{eqnarray}
\mathcal{J}(\mathbf{w}) = \min_{\mathbf{w} \in \mathbb{R}^{d}} \frac{1}{T}\sum_{t=1}^{T}\left( y_{t} - (\mathbf{x}_{t})^{\top}\mathbf{w} \right)^{2} + \lambda \lVert\mathbf{w} \rVert^{2}
\label{eq:7}
\end{eqnarray}
The data collection frequency is every 5 minutes over the time horizon of 100 days, resulting in a dataset size of $N= 9,044,683$, assembling roughly 250 values measured per ED and per day. All data is normalized and scaled, i.e., each parameter $x \in \mathbb{R}$ is mapped to $\frac{x-\mu}{\sigma}$ with mean value $\mu$ and variance $\sigma$ and scaled in the unity interval, 
thus, 
$\mathbf{x} \in [0,1]^{d}$.  

A training period of 1 month is considered before testing the proposed methods. 
This splitting of DS of size $N=N_T+N_M$ results in $N_T=1,500,000$ data points for the training period and leaves around $N_M=7,500,000$ data points for performing the proposed privacy-efficient approaches. Converting this into a percentage, only 15.8\% of the collected values are used during the training period to overcome the cold-start problem using transfer learning, and over 84\% is used for testing the different approaches. The starting date is located on the 1st of January 2018 and presents the time $t=s=0$. During the training period, each ED 
generates a local model $f_k$. At time $t=0$, each ED sends its local model $f_k$ towards the CL. Inside the CL, these models are merged using FedAVG as defined in Equation (\ref{eq:ch5_fedavg}) towards the first generalized central model $f_{FL}$. At time $t=s=0$, 
$f_{FL}$ is distributed to the EDs as 
the
first federated model. 


\subsection{Performance Metrics} We assess our methods with respect to two categories of performance metrics for accuracy and information loss. 

For the former, we are using three different metrics acknowledged in the literature: (1) Mean absolute error (MAE); (2) Root Mean Squared Error (RMSE) $= \frac{1}{N}\sum_{n=1}^{N}(\hat{y}_{n}-y_{n})^{2}]^{1/2}$; and (3) Symmetric mean absolute percentage error (SMAPE) \newline $=\frac{100}{T} \sum _{t=1}^{T}\frac{|\hat{y}_{t}-{y}_{t}|}{|\mathbf{y}_{t}|+|\hat{\mathbf{y}}_{t}|}$, because of its unbiased properties and ability to compare the results representing in percentage with values in $[0,100]$. 

Analyzing the information loss we use the metric of Kullback-Leibler (KL) divergence. The KL divergence from $p(\mathbf{x})$ to $p(\tilde{\mathbf{x}})$ denotes the information loss when attempting to reconstruct time series $\tilde{\mathbf{x}}$ for the actual time series $\mathbf{x}$, using $p(\tilde{\mathbf{x}})$ and $p(\mathbf{x})$ as the probability distribution functions, respectively. KL is defined as: 
\begin{eqnarray}
KL(p(\tilde{\mathbf{x}})\rVert p(\mathbf{x})) = \int_{0}^{1}p(\tilde{\mathbf{x}}) \log \frac{p(\tilde{\mathbf{x}})}{p(\mathbf{x})} dx.
\label{eq:ch3_kl}
\end{eqnarray} 

\subsection{Comparative Assessment}
\label{sec:com_ass}
A baseline for comparison is needed to evaluate the performance of the privacy-efficient analytics methodologies proposed in this paper. 
Therefore, during the assessment, the performance of the \textbf{Global Model (G)} with transmitting raw data to the CL from each ED at each time 
instance
$t$ is used as a comparison model towards the others. 
Concentrating on the privacy aspect of local 
ML models
and analytics, further methods are the basic deployment of FL represented during the assessment with \textbf{Federated Model (FM)}. The contrary is the other deployment of using just the local model \textbf{Local Model (L)} without any usage of the central coordinator or generalized model. The local model is built from stretch without any previous model coming from the CL.

Given these three baseline methods 
we compare the following introduced methodologies of personalized privacy-efficient learning: 

\begin{enumerate}
    \item The \textbf{Evolving Federated Model (EFM)} uses as initial local model $f_k$ the at time $t=0$ received generalized model $f_{FL}$. At each selected epoch $s$, the CL requests the local model and merges the generalized $f_{FL}$ over them. Only 
    $f_k$ is stored inside the ED.
    
    \item The \textbf{Local Federated Model (LFM)} is extending the EFM by updating the generalized model $f_{FL}$ locally at each time $t$ until the new update from the CL is sent at epoch $s$.
    
    \item The \textbf{Adaptive Selection Method (ASM)} introduces a dual parallel model implementation inside each ED. A local model $f_k$ is set to $f_k=f_{FL}$ at $s=t=0$ and is continuously updated each time $t$. The generalized model $f_{FL}$ is received and updated as the FM. The final prediction is generated through a reward function with respect to the historical performance of each model and balancing the $f_k$ and $f_{FL}$ model through the parameter $\alpha$.
    
    \item The \textbf{Smoothed Model (SM)} is based on the same concept as the ASM but with a fixed value for the balancing weight $\alpha$.
    
    \item The \textbf{Time-Optimized Selection Method (TOSM)} provides a selecting mechanism of the optimal model for this time. The fundamental concept is based on finding the optimal time $t^*$ to switch between the two models $f_k$ and $f_{FL}$ and vice versa using OST.
    
\end{enumerate}

The computational complexity of SM, ASM, LFM and EFM is for each ED $i$, $\mathcal{O}(d+1)$ each time $t$ and at the CL $\mathcal{O}(k+d+1)$ each time $s$.

\subsection{Parameters Configuration} 
Assessing the accuracy, multiple parameters for the different introduced methods and ED storage capacities have to be set. After the starting point at $t=0$ each time $t$, the models of EFM, LFM, L are updated using SGD. The values for $\theta$ are inserted into the SLW $\mathcal{O}$ each time $t$ for the proposed ASM strategy. The adaptive weighting $\alpha$ is generated through the ratio over this SLW $\mathcal{O}$ of size $U$. In the assessment $U$ is set to $U=\{50,100,250,500\}$. Moreover, for the TOSM method, the $Z_t$ and $Q_t$ values are defined each time $t$ and the respective current optimal model flagged inside each ED $k$. The delay tolerance level for the TOSM strategy is defined with $\beta$ and set to be $\beta = \{0.1,0.3,0.5,0.7,0.9\}$. If the CL requests at epoch $s$ an update of the FM, the values inside the SLW $\mathcal{W}$ are used for the SGD. The assessment parameters for the SLW $\mathcal{W}$ of size $M$ is set to be the previous day and the time since the last update was requested, so $M=\{250, 500, 1000\}$. The request epoch $s$ is set for the assessment to be every day, every other day and every fourth day so that $S=\{74,37,10\}$ and at the times $t=\{s\cdot250,s\cdot500,s\cdot1000\}$ respectively. 

Assessing the overall performance of each proposed method, a type of cross-validation has been deployed to guarantee independent validation of the results. The application is stopped at 24 random time points $t$. At the selected time $t$ the methodology implemented is stopped and the next $250$ values (representing one day) of each ED $k$ is used as prediction input to analyze the performance of the different privacy-efficient analytics strategies.

\section{Experimental Assessment}
\label{sec:expresults}

\subsection{Personalized Privacy-Efficient Learning}

\begin{figure}[h!]
    \centering
    \begin{subfigure}[b]{.49\columnwidth}
    \centering
    \includegraphics[width=1.0\linewidth]{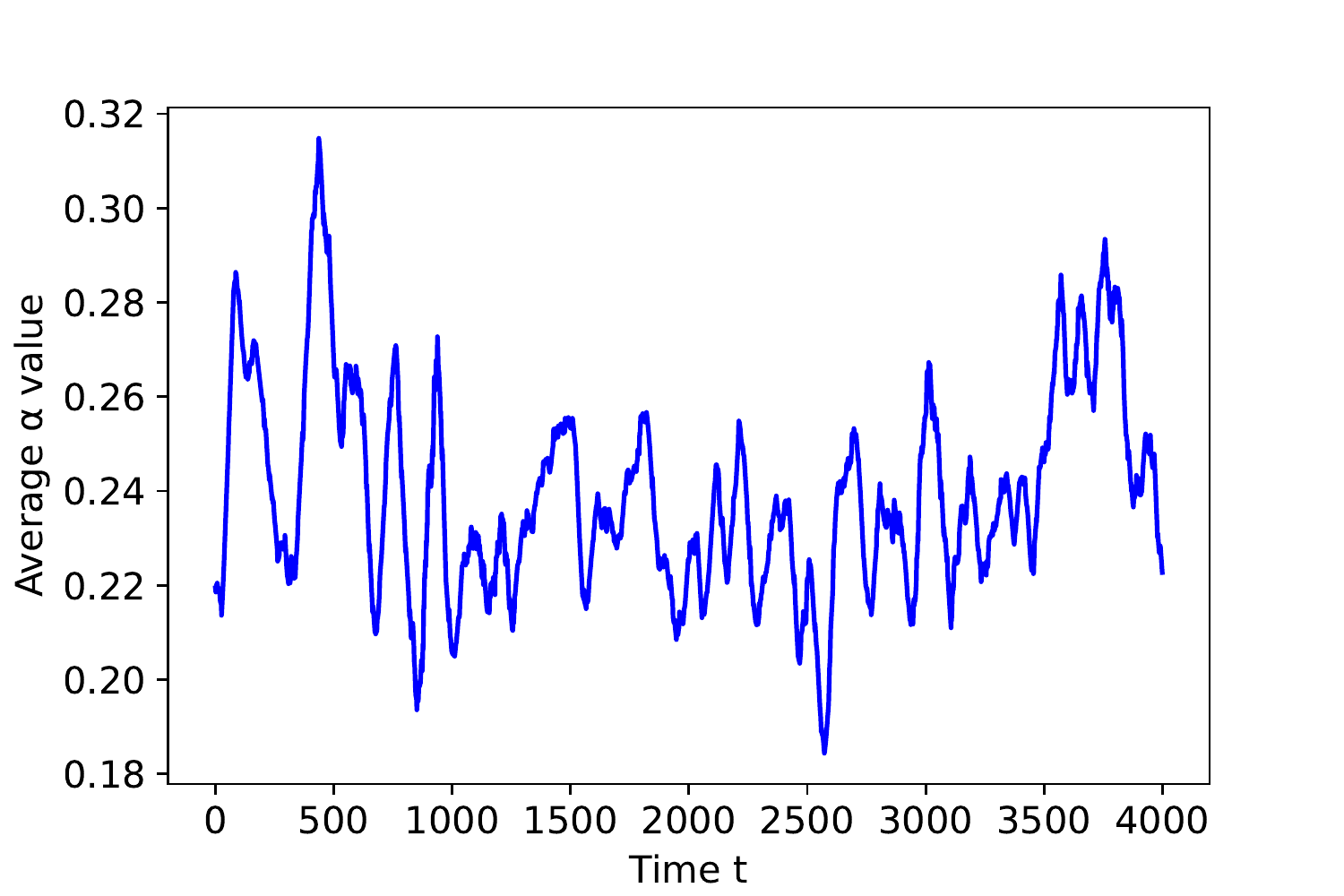}   
    \caption{Average value of $\alpha$ over time $t$}
    \end{subfigure}\hfill
    \begin{subfigure}[b]{.49\columnwidth}
    \centering
    \includegraphics[width=1.0\linewidth]{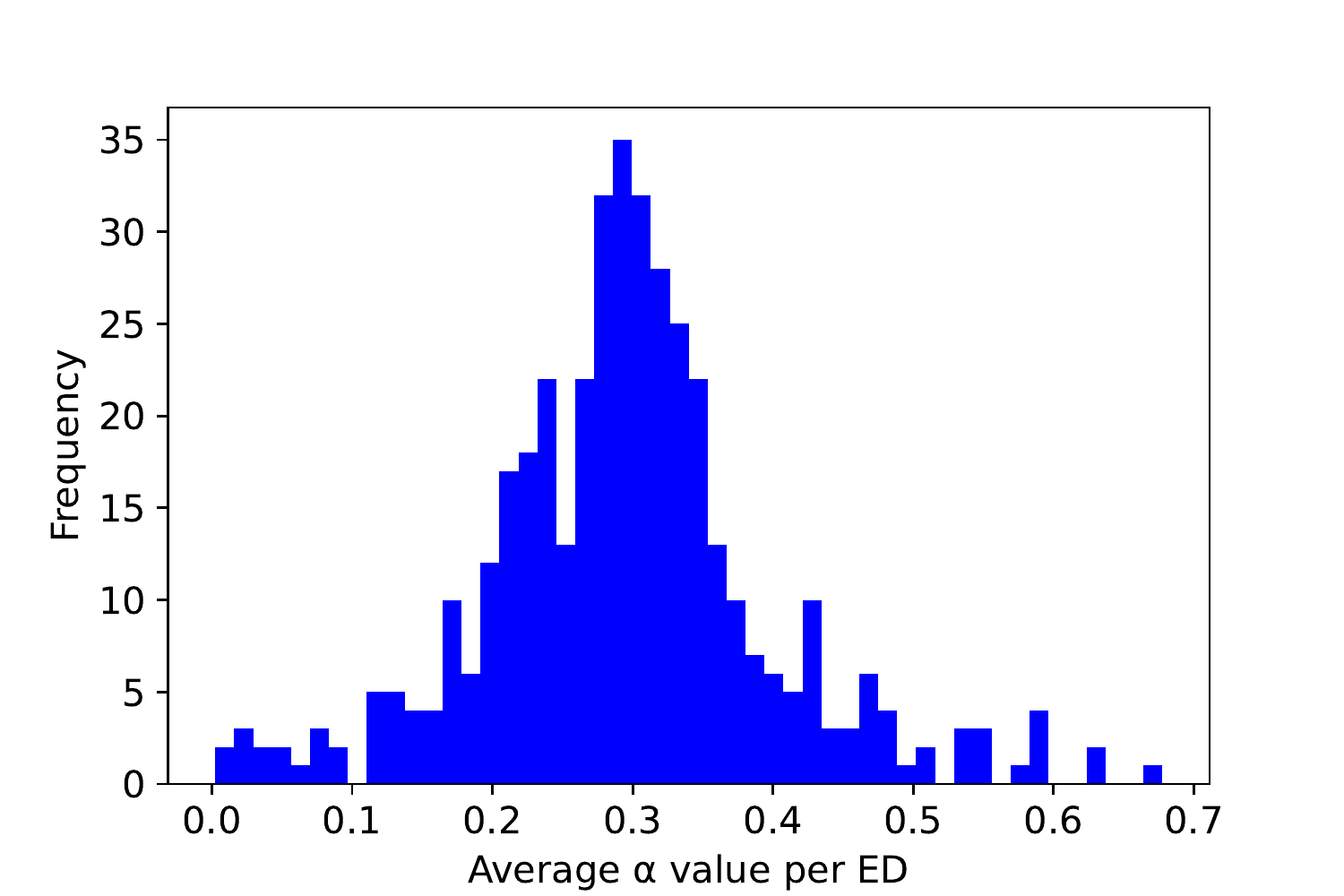}   
    \caption{Histogram of average $\alpha$ in EDs}
    \end{subfigure}
    \begin{subfigure}[b]{.49\columnwidth}
    \centering
    \includegraphics[width=1.0\linewidth]{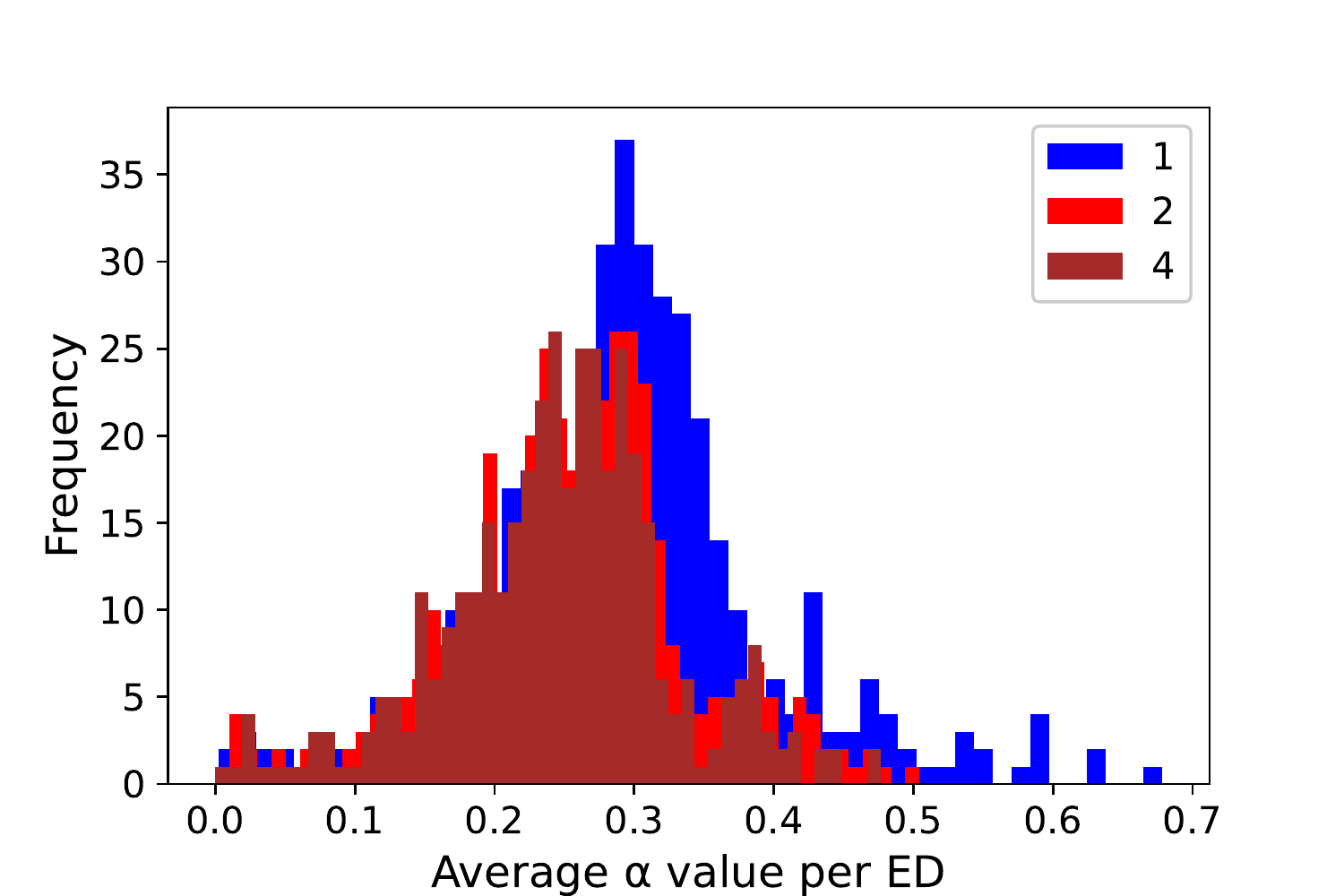}   
    \caption{Histogram of average $\alpha$ in EDs depending on the setting of the federated update frequency $s$}
    \end{subfigure}\hfill
    \begin{subfigure}[b]{.49\columnwidth}
    \centering
    \includegraphics[width=1.0\linewidth]{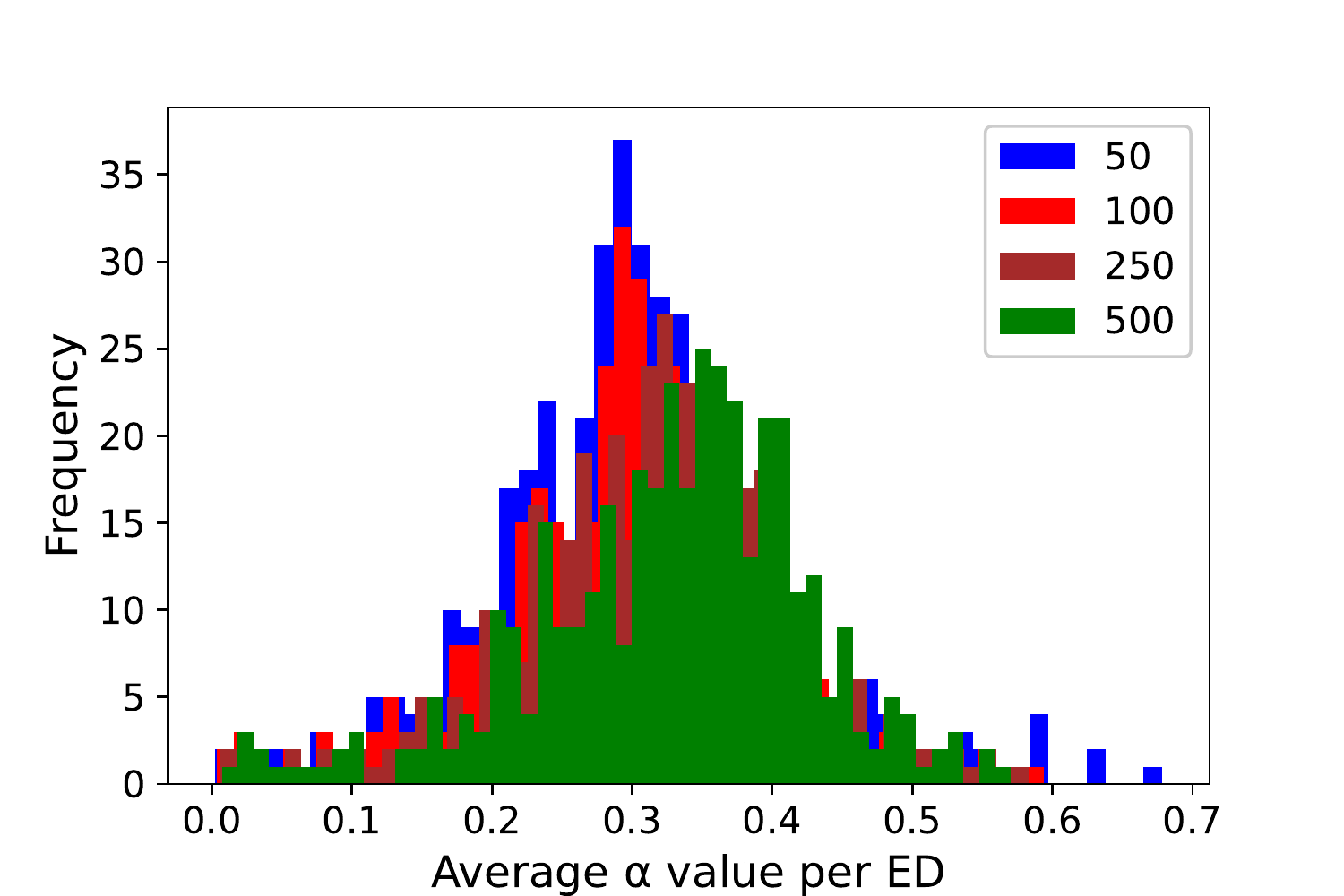}   
    \caption{Histogram of average $\alpha$ in EDs depending on the setting of SLW $\mathcal{O}$ of different length $U$}
    \end{subfigure}\hfill
    \caption{Parameter setting influence on $\alpha$ for the ASM }
    \label{fig:ch5_alpha}
\end{figure}

The assessment of the previously introduced methods is highly dependent on the parameter settings explained in . This section starts with analyzing the behavior of $\alpha$ indicating the weighting between local model prediction using $f_k$  and the generalized federated model $f_{FL}$ for the weighted prediction $\hat{y}$ using the proposed method of ASM.The influence of $s$ (epoch of CL request and updates the $f_{FL}$ model) and $U$ (SLW $\mathcal{O}$ size of considered rewards $\theta$) to the model weighting parameter $\alpha$ using the introduced performance metrics are illustrated in \autoref{fig:ch5_alpha}.

In \autoref{fig:ch5_alpha} (a) the average value of $\alpha$ over all EDs 
until $t=4000$ is presented. It can be seen that the value of $\alpha$ is greatly changing over time but lies in the range of $\alpha\geq0.18$ and $\alpha\leq0.32$. This figure has been using the setting of $U=50$, $M=1000$, and $s$ to be every day with $S=74$. The same settings are used for the frequency analysis and $\alpha$ distribution over each ED 
illustrated in \autoref{fig:ch5_alpha} (b). This figure shows the distribution of average $\alpha$ values for each ED and can be identified as a normal distribution with the mean around $\alpha=0.3$. This correlates with the findings of \autoref{fig:ch5_alpha} (a) for the average $\alpha$-values per time instance $t$ over the entire EDs $K$ lying in the highlighted range. To identify the influence of $s$ and $U$ towards the weighting parameter $\alpha$, \autoref{fig:ch5_alpha} (c)  and \autoref{fig:ch5_alpha} (d) highlight this respectively. In \autoref{fig:ch5_alpha} (c), the setting of $s=\{1,2,4\}$ indicating the update frequency of the federated model $f_{FL}$ to be each day, every second day, and every fourth day. From this figure, the influence of the update frequency $s$ towards the model weighting $\alpha$ indicated that increasing the frequency is decreasing the mean of $\alpha$ and increases the variance. In opposite to the influence of $U$ to $\alpha$, illustrated in \autoref{fig:ch5_alpha} (d). In this figure, an increase of the mean is presented by increasing the SLW size of rewards to be considered. The rise of $U$ is only influencing the mean but not the variance for the value $\alpha$ in each ED $k$.

\begin{figure}[h!]
    \centering
    \begin{subfigure}[b]{.49\columnwidth}
    \centering
    \includegraphics[width=1.0\linewidth]{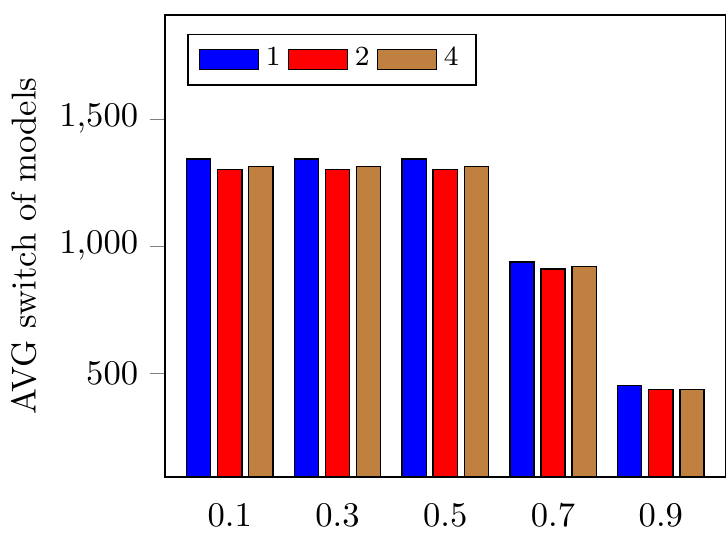}%
    \caption{Average switching time for different $\beta$ and federated update frequency $s$}
    \end{subfigure}\hfill
    \begin{subfigure}[b]{.49\columnwidth}
    \centering
    \includegraphics[width=1.0\linewidth]{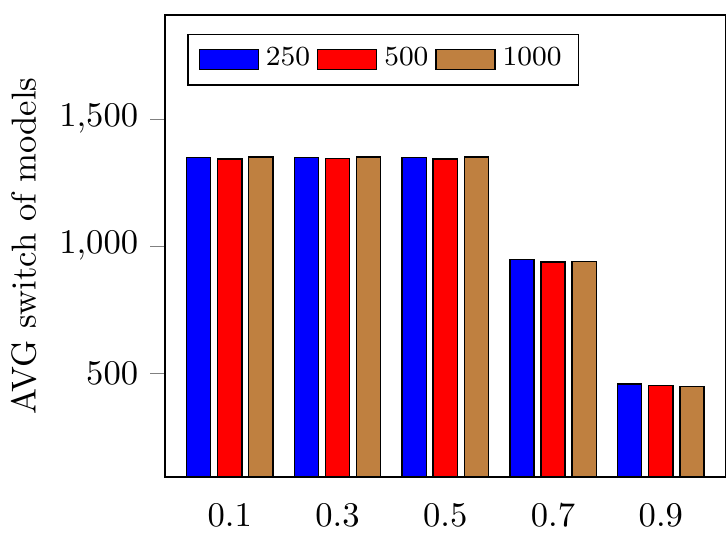}%
    \caption{Average switching time for different $\beta$ and SLW sizes for $M$ \newline}
    \end{subfigure}\hfill
    \begin{subfigure}[b]{.49\columnwidth}
    \centering
    \includegraphics[width=1.0\linewidth]{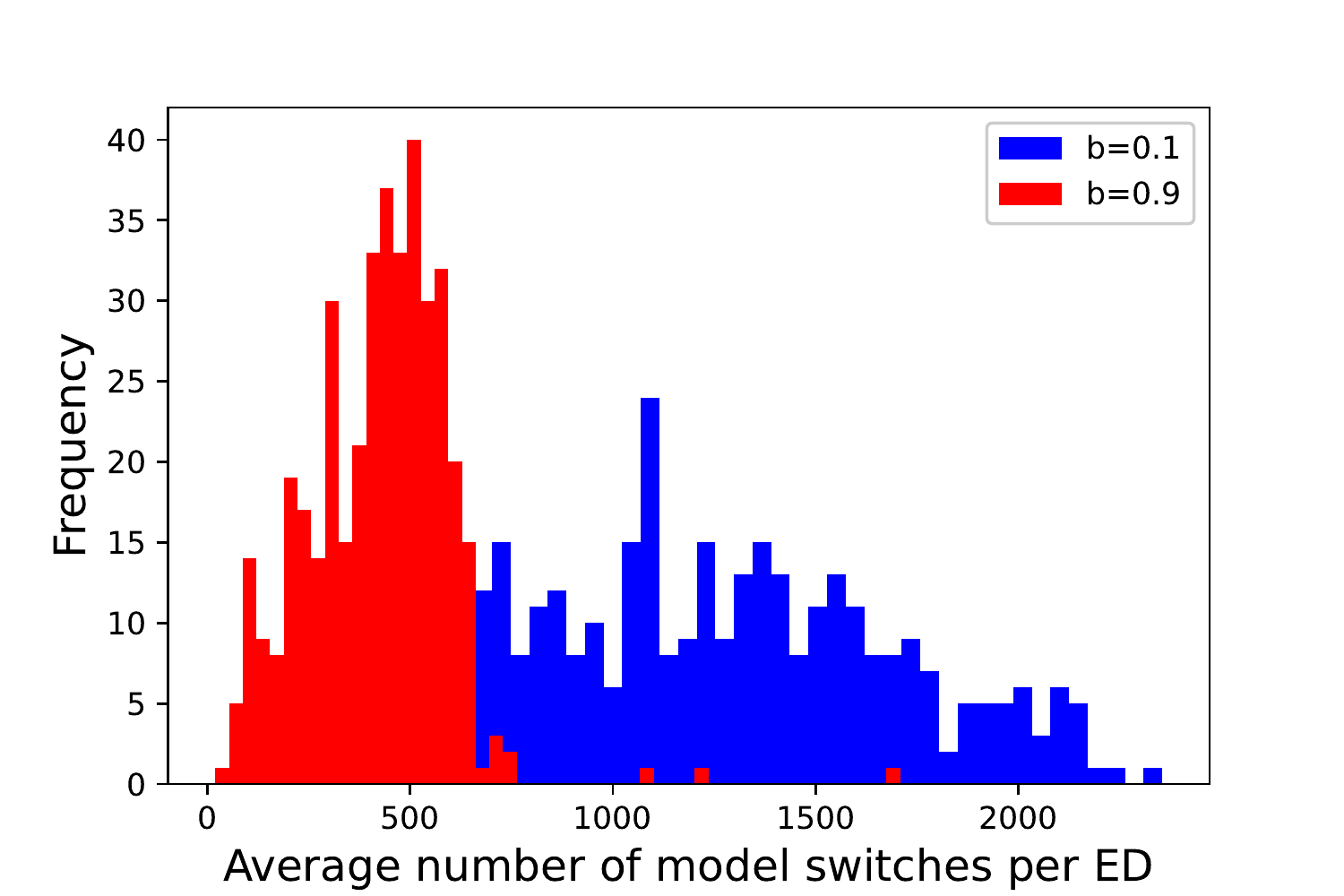}%
    \caption{Histogram of average switching time for \\ $\beta=\{0.1,0.9\}$ using $M=250$}
    \end{subfigure}\hfill
    \begin{subfigure}[b]{.49\columnwidth}
    \centering
    \includegraphics[width=1.0\linewidth]{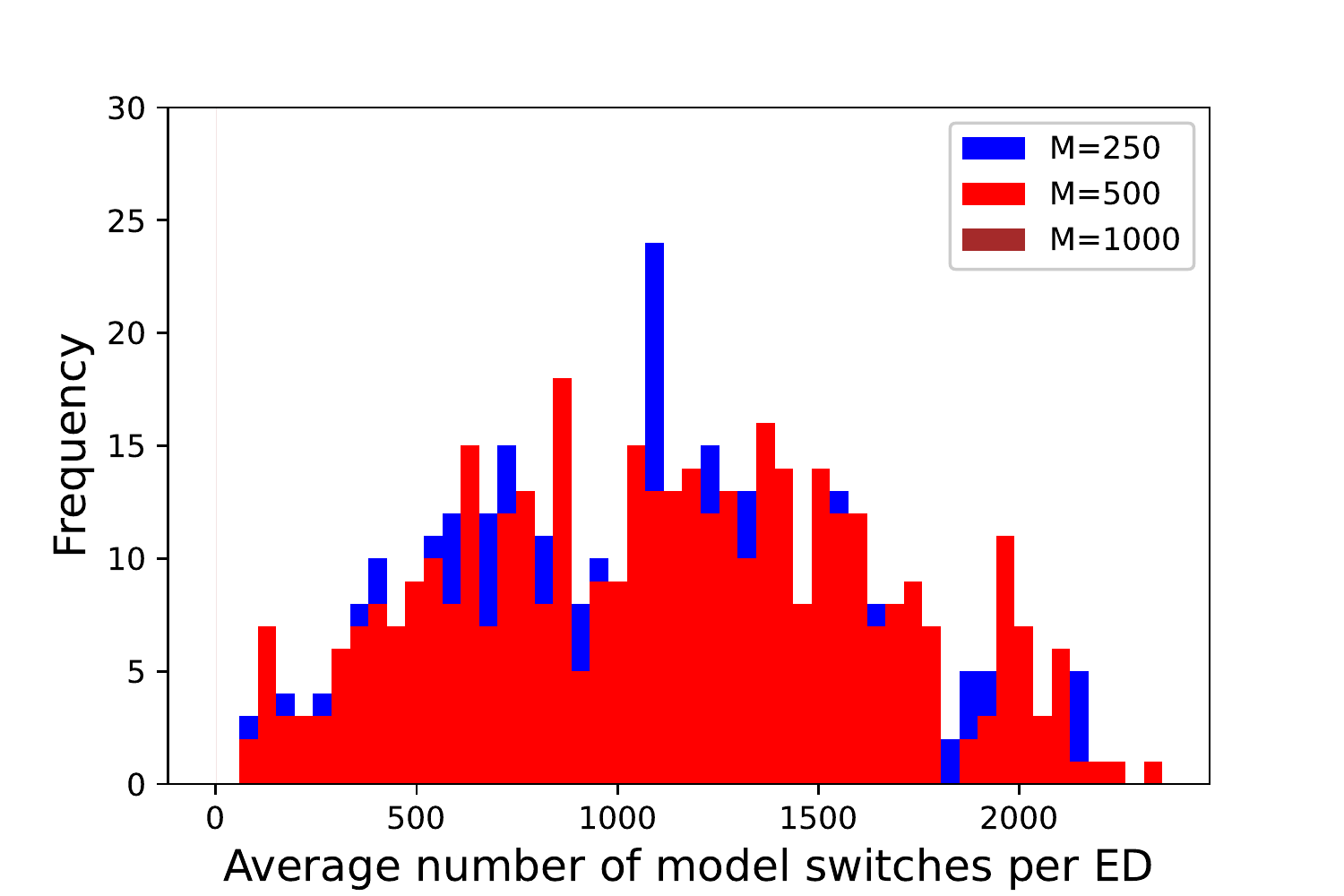}%
    \caption{Histogram of average switching time for $\beta=\{0.5\}$ using $M=\{250, 500, 1000\}$}
    \end{subfigure}\hfill
    \caption{Parameter setting influence on $\beta$ for the TOSM}
    \label{fig:ch5_beta}
\end{figure}

The next proposed model in which the parameters have to be first analyzed is the TOSM. Introducing the value of $\beta$ for the delay tolerance of switching the models $f_k$ and $f_{FL}$ to use the optimal model at the prediction time. This value of $\beta$ is analyzed in \autoref{fig:ch5_beta} with the number of model switches that occur during the runtime. In \autoref{fig:ch5_beta} (a), the influence of switching the models dependent on $\beta$ with increasing the update frequency of $s$ is illustrated. In this figure, no difference between the variation of $s$ can be seen influencing the model switching of the TOSM strategy. However, a decrease of average switches can be seen when the value of $\beta \geq 0.7$. Similar results are highlighted in \autoref{fig:ch5_beta} (b). This illustration shows the influence of increasing the SLW $\mathcal{W}$ size $M$ with respect to $\beta$ and the number of times the model is switched inside the ED. \autoref{fig:ch5_beta} (c) shows the distribution of $\beta=0.1$ and the number of switches inside each ED and the distribution of $\beta=0.9$ with delaying the switching. In $\beta=0.9$, the variation of average switches is higher than with $\beta=0.1$ in which the density is around the mean of 500. In \autoref{fig:ch5_beta} (d), the distributions of $\beta=0.1$ using different values of $M$ are presented. Already shown in \autoref{fig:ch5_beta} (b), no change of the frequency by differing the size of $M$ can be seen.

\begin{figure}[h!]
    \centering
    \begin{subfigure}[b]{.48\columnwidth}
    \centering
    \includegraphics[width=1.0\linewidth]{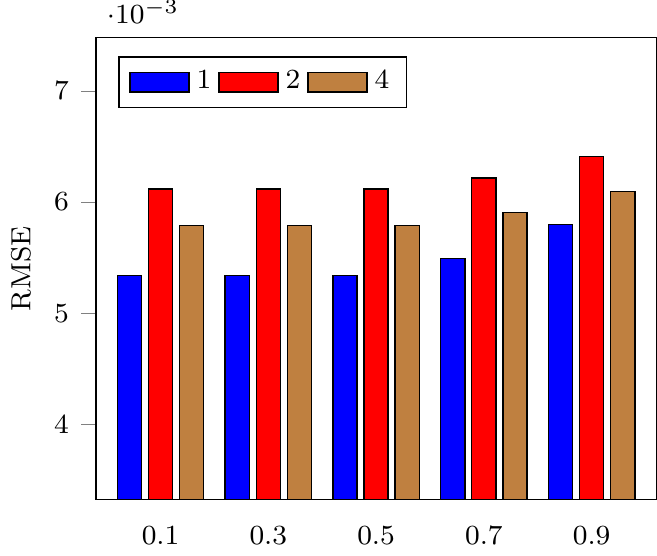}%
    \caption{Average RMSE for different $\beta$ and update  frequency $s$}
    \end{subfigure}\hfill
    \begin{subfigure}[b]{.49\columnwidth}
    \centering
    \includegraphics[width=1.0\linewidth]{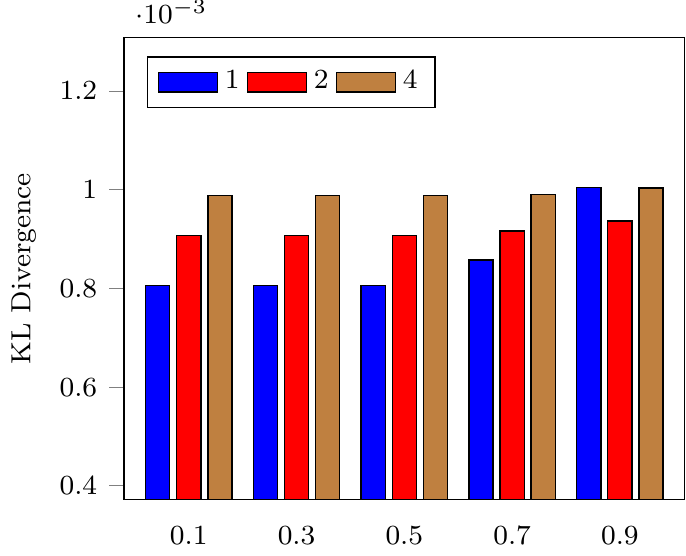}%
    \caption{Average KL for different $\beta$ and update frequency $s$}
    \end{subfigure}
    \begin{subfigure}[b]{.48\columnwidth}
    \centering
    \includegraphics[width=1.0\linewidth]{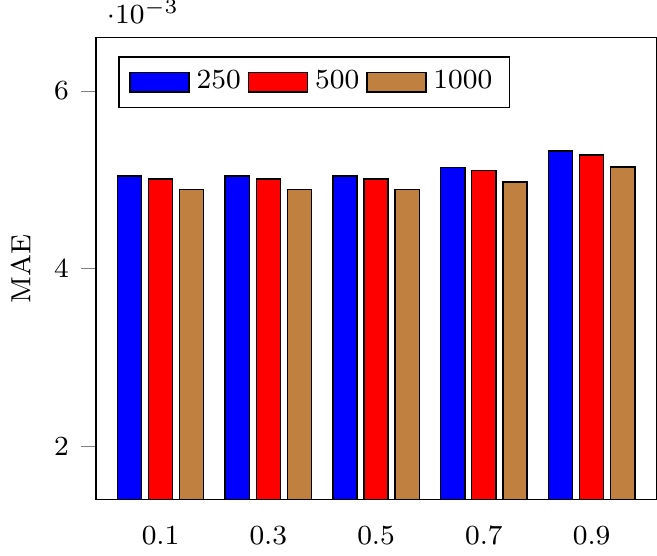}%
    \caption{Average MAE for different $\beta$ and SLW sizes for $M$}
    \end{subfigure}%
    \begin{subfigure}[b]{.49\columnwidth}
    \centering
    \includegraphics[width=1.0\linewidth]{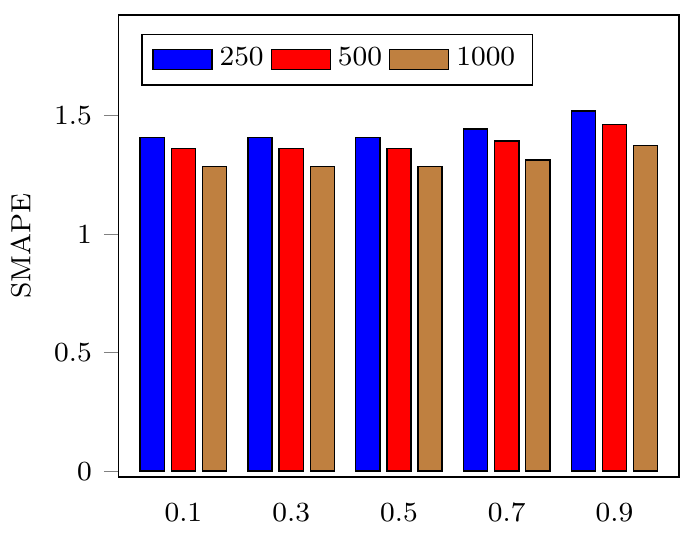}%
    \caption{Average SMAPE for different $\beta$ and SLW sizes for $M$}
    \end{subfigure}
    \caption{Comparison of the influence of delay tolerance $\beta$ for TOSM using different $s$ and $M$ towards the performance metrics KL, RMSE, MAE, and SMAPE}
    \label{fig:ch5_beta_accuracy}
\end{figure}

In \autoref{fig:ch5_beta_accuracy}, the performance for RMSE, MAE, SMAPE and KL divergence over the different possible $\beta$ values is analyzed. Highlighted in the previous figure, increasing the delay of switching between the models $f_k$ and $f_{FL}$ inside the ED indicated through the value of $\beta$ shows that $\beta \rightarrow 1$ increases the tolerance and less often the model is switched.

In \autoref{fig:ch5_beta_accuracy} (a), the analysis of the performance of RMSE with respect to the update frequency $s$ is illustrated. It can be seen that the update frequency of $s$ is only slightly increasing the error represented by the metric RMSE. Similar results have been generated using MAE and SMAPE, but these results are removed to space limitations. Moreover, in this figure, the behavior of $s$ towards the accuracy is illustrated. Using $s=1$ indicating an update frequency of the $f_{FL}$ model every day and results in the lowest accuracy error. However, assuming that increasing the frequency is rising the error does not hold as the performance of $s=4$ (update frequency every fourth day) generates lower prediction errors than $s=2$. The information loss metric KL divergence highlighted in \autoref{fig:ch5_beta_accuracy} (b) indicate a clear decrease of information loss by increasing the update frequency $s$. Additionally, the influence of $\beta$ towards the information gain shows that using more frequent switching of models results in higher entropy, presented by the KL metric, inside the ED. 

\autoref{fig:ch5_beta_accuracy} (c) and \autoref{fig:ch5_beta_accuracy} (d) investigate the behavior of increasing the SLW $\mathcal{W}$ of size $M$ representing the data used for performing SGD on at time $t=s$ for updating the $f_{FL}$ to the CL. In \autoref{fig:ch5_beta_accuracy} (c), the influence is presented by the metric MAE. Increasing the size $M$ to $M=1000$ is decreasing the error. The dependency of $\beta$ with respect to the accuracy over different SLW sizes $M$ does show a slight increase of the MAE when increasing the delay tolerance of $\beta \rightarrow 1$. In \autoref{fig:ch5_beta_accuracy} (d), similar behavior and dependency are illustrated. Raising $\beta \rightarrow1 $ is increasing the SMAPE and increasing the SLW $\mathcal{W}$ of size $M$ to $M=1000$ is decreasing the prediction error. 

\begin{figure}[h!]
    \centering
    \begin{subfigure}[b]{.49\columnwidth}
    \centering
    \includegraphics[width=1.0\linewidth]{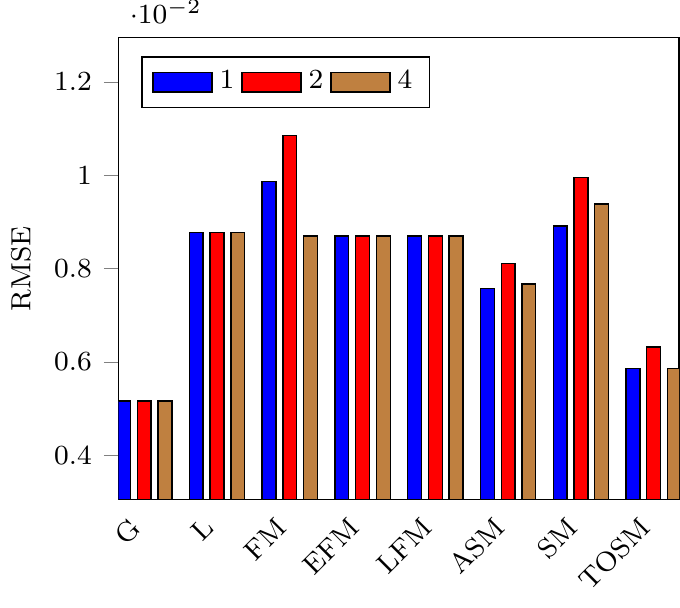}%
    \caption{Performance for RMSE divergence using \\ $U=250$ and $M=250$}
    \end{subfigure}\hfill
    \begin{subfigure}[b]{.49\columnwidth}
    \centering
    \includegraphics[width=1.0\linewidth]{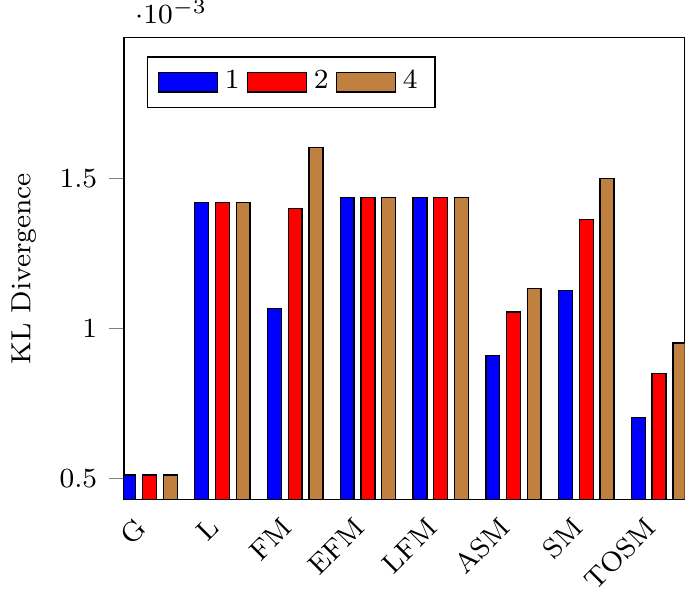}%
    \caption{Performance for KL divergence using $U=50$ and $M=1000$}
    \end{subfigure}
    \begin{subfigure}[b]{.48\columnwidth}
    \centering
    \includegraphics[width=1.0\linewidth]{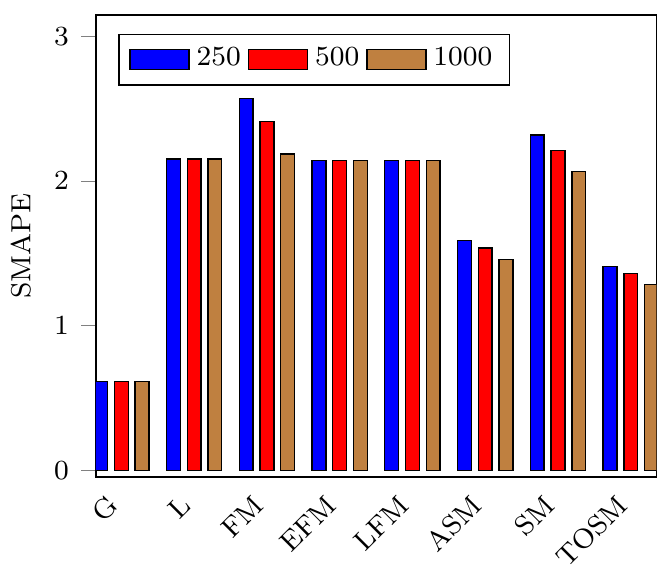}%
    \caption{Performance for SMAPE using $U=50$  and $s=1$}
    \end{subfigure}\hfill
    \begin{subfigure}[b]{.49\columnwidth}
    \centering
    \includegraphics[width=1.0\linewidth]{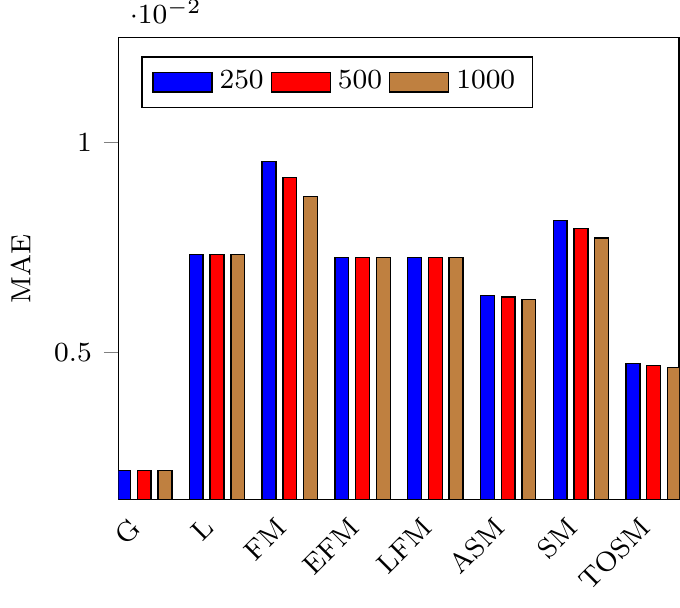}%
    \caption{Performance for MAE using $U=250$ and $s=4$}
    \end{subfigure}\hfill
    \caption{Performance evaluation over all proposed models using different settings for $s$ and $M$ using the performance metrics KL, RMSE, MAE, and SMAPE}
    \label{fig:ch5_s_m_influence}
\end{figure}

In \autoref{fig:ch5_s_m_influence}, the performance of all models introduced in \autoref{sec:com_ass} and the influence of the parameters $s$ and $M$ is illustrated for assessment. In \autoref{fig:ch5_s_m_influence} (a), the metric of RMSE for $M=250$ and $U=250$ over all epoch selection variables $s$ is highlighted. In this figure the influence of $s$ towards the models FM, ASM, SM and TOSM is illustrated. The models GM, L, EFM, and LFM do not change their performance by changing the parameters of $M$ or $s$ as their learning and adaptation to the data input is continuously and independent of the FM. Increasing the frequency of $s$ is increasing the prediction error for the method SM and ASM. Whereas the accuracy is increasing for the FM method. The TOSM approach is as indicated in \autoref{fig:ch5_beta_accuracy} only slightly increasing and performing worst for $s=2$, representing updating every other day. Further, this figure shows that the GM generates the best accuracy, but the TOSM strategy provides accuracy closest to the GM. In \autoref{fig:ch5_s_m_influence} (b) the information loss over changing update frequencies $s$ over all models is highlighted. Using the setting of $M=1000$ and $U=50$ with the performance metric KL divergence, the results show that for the models depending on these parameters (FM, ASM, SM and TOSM ) the KL is increasing with the increase of the update frequency $s$. A greater information loss can be seen for all four affected methods comparing $s=1$ and $s=4$ with each other. Moreover, the information loss is smallest when transmitting raw data and generating a centralized model (GM). 
\autoref{fig:ch5_s_m_influence} (c) and \autoref{fig:ch5_s_m_influence} (d) highlight the performance towards changing the SLW $\mathcal{W}$ size $M$ and the corresponding behavior of each model. Setting the parameters for \autoref{fig:ch5_s_m_influence} (c) with $s=1$ and the SLW $\mathcal{O}$ size $U=50$ using the metric SMAPE shows the following results. Over all dependent models, the increase of the training dataset size in the SLW $\mathcal{W}$ inside each ED $k$ for SGD presented by $M$ is decreasing the accuracy error. In \autoref{fig:ch5_s_m_influence} (d), similar behavior is illustrated with the update frequency of $s=4$ and $U=250$ using the MAE performance metric. 

\begin{figure}[h!]
    \centering
    \begin{subfigure}[b]{.49\columnwidth}
    \centering
    \includegraphics[width=1.0\linewidth]{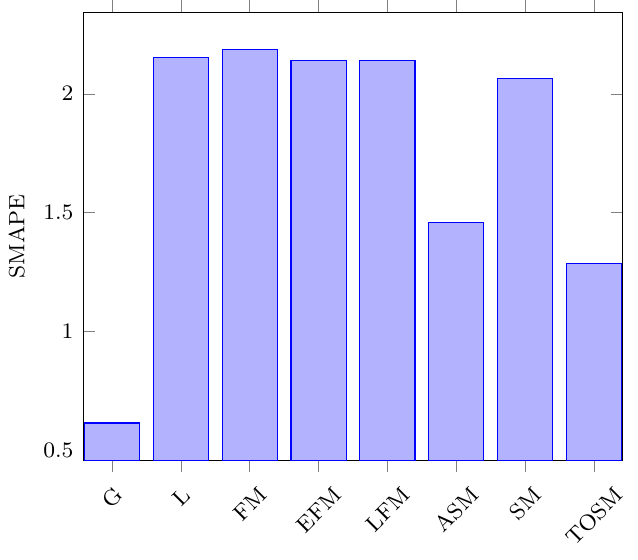}
    \caption{Performance with best parameter setting on SMAPE}
    \end{subfigure}\hfill %
    \begin{subfigure}[b]{.48\columnwidth}
    \centering
    \includegraphics[width=1.0\linewidth]{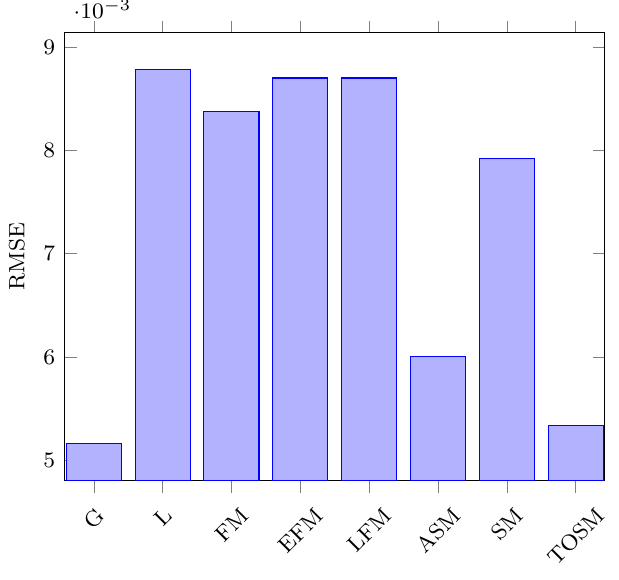}
    \caption{Performance with best parameter setting on RMSE}
    \end{subfigure}
    \begin{subfigure}[b]{.48\columnwidth}
    \centering
    \includegraphics[width=1.0\linewidth]{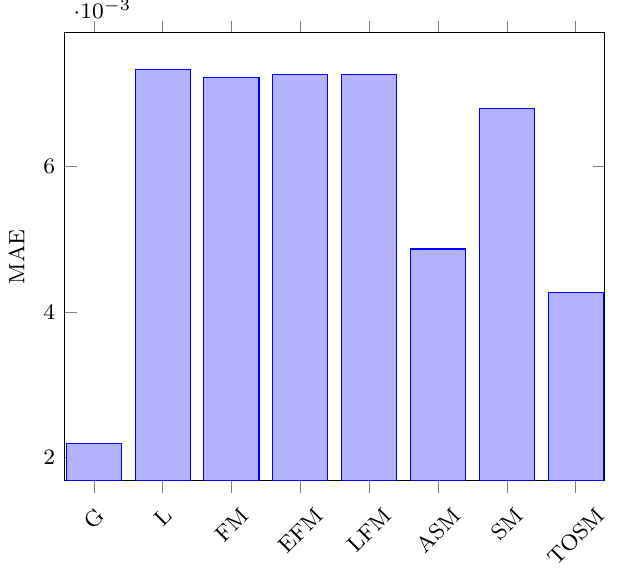}
    \caption{Performance with best parameter setting on MAE}
    \end{subfigure}\hfill %
    \begin{subfigure}[b]{.49\columnwidth}
    \centering
    \includegraphics[width=1.0\linewidth]{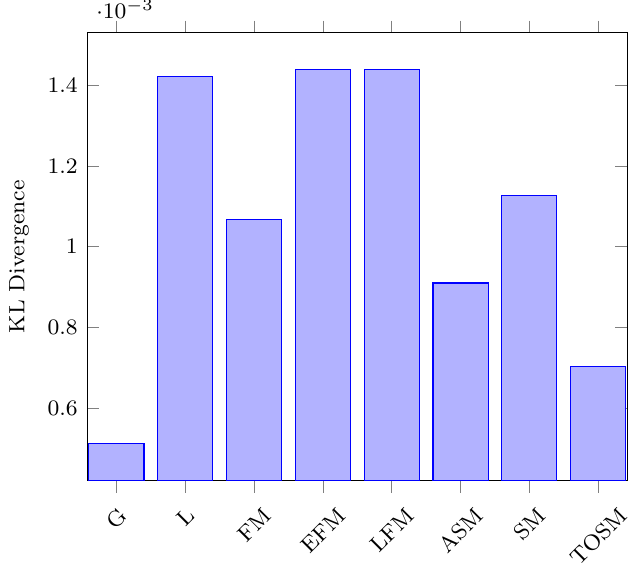}
    \caption{Performance with best parameter setting on KL divergence}
    \end{subfigure}\hfill
    \caption{Comparison of using the best parameter settings towards the performance metrics KL, RMSE, MAE, and SMAPE}
    \label{fig:ch5_best_off}
\end{figure}

After identifying the influence of the SLW $\mathcal{W}$ of size $M$, the SLW $\mathcal{O}$ of length $U$ for the ASM approach, and the central defined update frequency $s$, towards the models proposed in this chapter (see \autoref{sec:com_ass}) with respect to their performance the following best settings have been used over the four performance metrics KL divergence, RMSE, MAE, and SMAPE and illustrated in \autoref{fig:ch5_best_off}. \autoref{fig:ch5_best_off} uses the setting of $M=1000$ as shown in \autoref{fig:ch5_s_m_influence} performing lowest of prediction error over all models, $U=50$ as highlighted in \autoref{fig:ch5_alpha} for minor variance in the average $\alpha$ value, $s=1$ indicating an update of the FM of each day with $S=74$, and $\beta=0.3$ analyzed in \autoref{fig:ch5_beta_accuracy} and \autoref{fig:ch5_beta} to be the most accurate value with respect to the used metrics. 

In \autoref{fig:ch5_best_off} (a), the metric of SMAPE over all assessed methods is compared using the best settings. From this figure, the aim is to identify what approach is closest to the GM with respect to their performance. Even though the GM is in privacy-preserving IoT applications not feasible, the aim is to have quality-aware and efficient models inside the ED performing as accurate as of the GM. Having this aspect in mind, the strategy of TOSM using the OST to identify the optimal time to switch between the local generated model $f_k$ (L) and the generalized model $f_{FL}$ (FM) shows the best performance. However, the introduced method of TOSM still increases the error by 1\%. The second best with regards to the performance metric SMAPE is the approach of ASM. The other methods do not show a great difference. \autoref{fig:ch5_best_off} (b) does the analysis using the metric of RMSE. In this figure, the proposed approaches differ more from each other. Highlighted in this figure is the great performance of TOSM with RMSE error very close to the GM. Moreover, the prediction accuracy performance of the FM is better than that of the L. ASM provides next to TOSM the best accuracy for predictive analytics in privacy-preserving environments. \autoref{fig:ch5_best_off} (c) and \autoref{fig:ch5_best_off}(d) illustrate these findings for the performance metrics MAE and KL divergence . Especially, for the presented KL divergence metric showing the information loss inside the ED by deploying the proposed mechanism, offer a great improvement to the FM by deploying a dual model strategy inside the ED and using either adaptive weighting with ASM or optimal selecting using TOSM.

\subsection{Behavior on Concept Drifts}

\begin{figure}[h!]
    \centering
    \begin{subfigure}[b]{.48\columnwidth}
    \centering
    \includegraphics[width=1.0\linewidth]{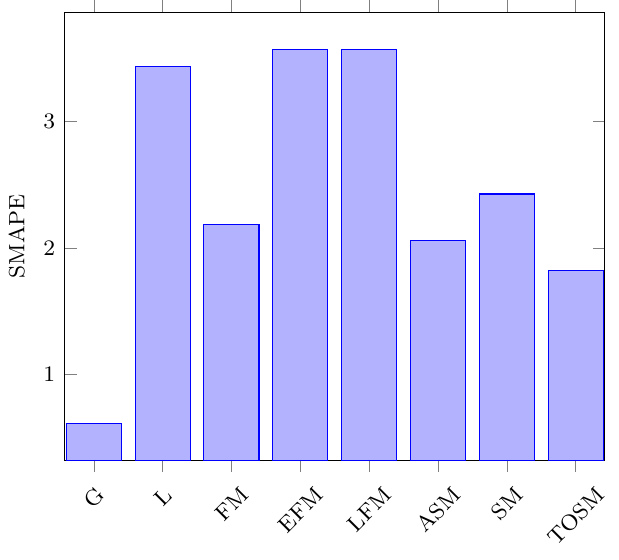}
    \caption{Performance on SMAPE}
    \end{subfigure}\hfill %
    \begin{subfigure}[b]{.49\columnwidth}
    \centering
    \includegraphics[width=1.0\linewidth]{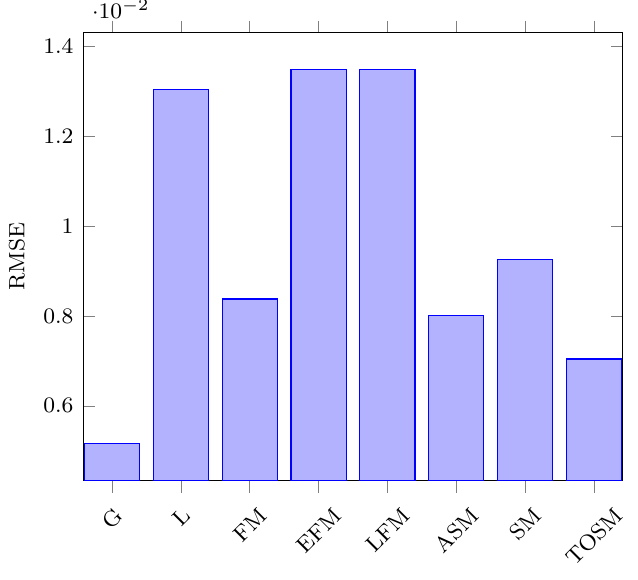}
    \caption{Performance on RMSE}
    \end{subfigure}
    \begin{subfigure}[b]{.49\columnwidth}
    \centering
    \includegraphics[width=1.0\linewidth]{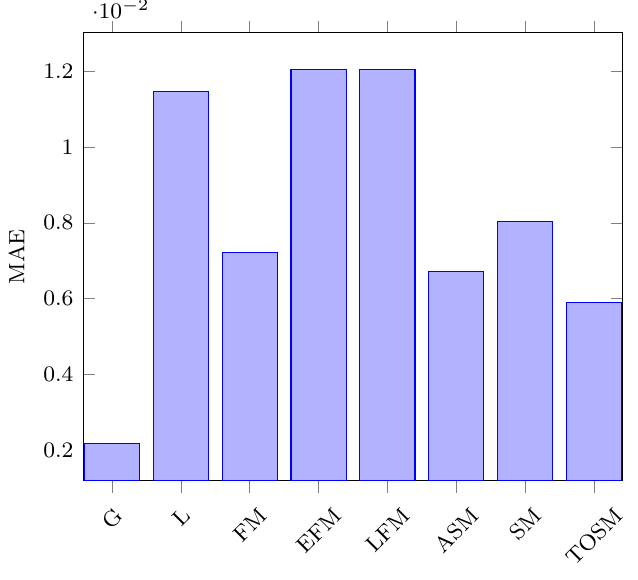}
    \caption{Performance on MAE}
    \end{subfigure}\hfill%
    \begin{subfigure}[b]{.49\columnwidth}
    \centering
    \includegraphics[width=1.0\linewidth]{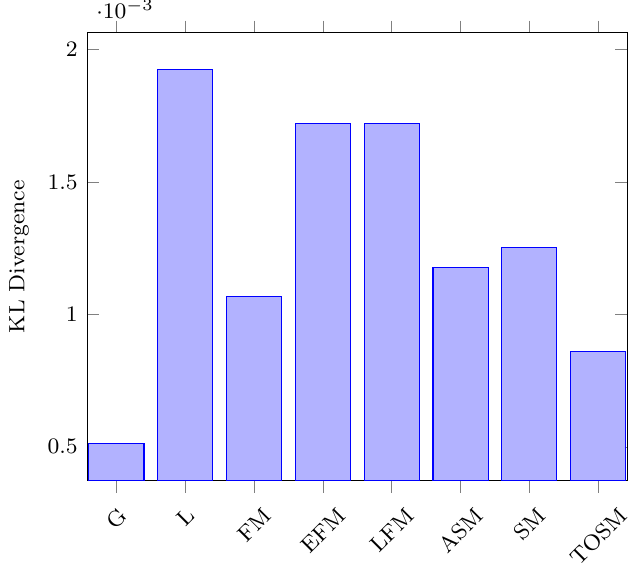}
    \caption{Performance on KL}
    \end{subfigure}\hfill
    \caption{Comparison of using the best parameter settings and artificially construct a concept drift towards the performance metrics KL, RMSE, MAE, and SMAPE}
    \label{fig:ch5_concept drift}
\end{figure}

This chapter argues the importance of deploying adaptive and evolving learning inside EDs to support qualitative predictive analytics and modelling. Assessing this hypothesis, the following section is introducing an artificially constructed concept drift to test the ability of each proposed method towards changing environments and quality-aware predictions. In \autoref{fig:ch5_concept drift}, the performance over all four metrics  KL divergence, RMSE, MAE, and SMAPE is illustrated using the identified best settings of \autoref{fig:ch5_best_off}. The values are compared against each other, showing the change of accuracy and information loss with concept drift appearance towards regular predictive tasks.

\autoref{fig:ch5_concept drift} (a) illustrates the performance with respect to the metric SMAPE. The illustration shows a clear improvement of the FM performance when  concept drift occurs. However, the adaptive and parallel model adaption of ASM and TOSM perform equal or better than the FM. This indicates the ability to adapt to changing environments when using the generalized model and incorporating the local individualized model towards a prediction. Moreover, it should be noted that the models of L, EFM, and LFM highly increase their error.  In \autoref{fig:ch5_concept drift} (c), the same behavior towards the approaches is illustrated using the MAE performance assessment metric. \autoref{fig:ch5_concept drift} (b) shows the RMSE during the concept drift appearance and highlights the good adaptation of ASM with similar performance to the presented values in \autoref{fig:ch5_best_off} (b) for familiar data inputs inside the ED $k$. The FM and ASM generate similar prediction results that indicate that the adaptive parameter $\alpha$ places more importance on the FM as the L can not adapt that fast to concept drifts. \autoref{fig:ch5_concept drift} (d) provides insights into the information loss by presenting the performance metric KL divergence.  The value of the KL divergence of the FM approach is improving through the concept drift, showing the importance of generalization inside EDs. Some improvement of the information loss value is given for the ASM method also, as it highly depends on the accuracy of the FM or LFM (depending on the weighting factor $\alpha$). However, the method of TOSM provides constantly low information loss independent of the occurrence of the concept drift or not. 

\section{Conclusion \& Future Work}

The focus on privacy-efficient analytics in resource constraint environments has been investigated. It has been shown through related work on privacy-preserving local edge learning, that Federated Learning introduced local learning and privacy of data by design. The newly developing research community leaves open research questions towards the quality and efficiency of Federated Learning under changing environments. In this paper, two strategies have been proposed that enable the ability to centrally learn a predictive model and enhancing the quality of local inferred and predictive results. Quality of analytical results through enabling the local individuality of heterogeneous devices provided the fundamentals of these approaches. The first model, Adaptive Selection Model, uses the local model and generalized model to provide a new prediction outcome by weighting these two models based on historical rewards. The second strategy introduces the optimization to find the best time to switch between the local model and the generalized federated model by using Optimal Stopping Theory. 
Through real data evaluating the effectiveness of both methods have been shown. Further, it was possible to provide evidence, that a mixture or switching strategy between models inside Edge Devices enables qualitative privacy-preserving predictive analytics for continuous changing and evolving environments (including concept drifts).

Based on the provided work in this paper further research can be conducted in hierarchical structuring different levels of Federated Learning models to enable not only personalized models at each ED but enable group-based models for similarities. Moreover, the combination of active learning for continuous changing environments and Federated Learning is highly interesting to combine and provide qualitative analytics for privacy-preserving and real-time critical environments over semi-supervised learning. 

\section*{Authorship Contribution Statement}
\textbf{Natascha Harth:} Conceptualization, Methodology,  Formal analysis, Writing- original draft, Visualization, Investigation, Software, Validation.
\textbf{Christos Anagnostopoulos:} Conceptualization, Formal analysis, Writing - review \& editing, Supervision.
\textbf{Kostas Kolomvatsos:}Writing - review \& editing, Supervision.
\textbf{Hans-Joerg Voegel:} Writing - review \& editing, Supervision.

\bibliographystyle{plain}


\end{document}